\documentclass[a4paper,11pt]{article}
\usepackage{jheppub}
\usepackage{amsfonts,placeins,pbox,multirow,mathtools}
\usepackage{rotate,xcolor,slashed,epstopdf,verbatim,url,multirow,booktabs}
\usepackage{cellspace,tabularx, comment,bbold}%
\usepackage{comment}
\usepackage{subfig}
\usepackage{amsmath}
\usepackage{bm}
\usepackage{array}
\usepackage{hyperref}
\usepackage[capitalise]{cleveref}
\usepackage{orcidlink}
\DeclareUnicodeCharacter{2212}{\textminus}

\renewcommand{\arraystretch}{1.5}

\numberwithin{equation}{section}

\newcommand{\tr}[1]{\left\langle {#1} \right\rangle}
\newcommand{\ChPT}{$\chi$PT}
\newcommand{\Order}[1]{\mathcal{O}\left( {#1} \right)}
\newcommand{\iu}{{i\mkern1mu}}
\newcommand{\Oph}{\mathcal{O}^h}
\newcommand{\Ophh}{\mathcal{O}^{hh}}
\newcommand{\Opl}{\mathcal{O}^{l}}
\newcommand{\State}[1]{\big| {#1}\big\rangle }
\newcommand{\BState}[1]{\big\langle {#1}\big| }

\newcommand{\BR}{\text{BR} }

\DeclareMathOperator{\darrowD}{\overleftrightarrow{\partial}_{\!\!\mu}}
\DeclareMathOperator{\darrowU}{\overleftrightarrow{\partial}^{\!\!\mu}}

\crefname{section}{Sec.}{Secs.}
\crefname{table}{Table}{Tables}
\crefname{figure}{Fig.}{Figs.}
\crefname{equation}{Eq.}{Eqs.}
\crefname{appendix}{Appendix}{Appendix}

\title{A Phenomenological Model of Mesons for Charged Current Weak Decays}
\author[a]{Sabyasachi Chakraborty\orcidlink{0000-0001-5356-7607}, }
\emailAdd{sabyac@iitk.ac.in}
\affiliation[a]{Department of Physics, Indian Institute of Technology, Kanpur 208016, India.}

\author[b]{Namit Mahajan\orcidlink{0009-0003-4655-6389},}
\emailAdd{nmahajan@prl.res.in}
\affiliation[b]{Theoretical Physics Division, Physical Research Laboratory, Ahmedabad 380009, India.}

\author[c]{and Tuhin S. Roy\orcidlink{0000-0002-1291-426X}}
\emailAdd{tuhin@theory.tifr.res.in}
\affiliation[c]{Department of Theoretical Physics, Tata Institute of Fundamental Research, Mumbai 400005, India.}

\date{}
\begin{document}

\abstract{
We propose a phenomenological model of pseudo scalar mesons to describe charged-current weak decays of heavy-light mesons. The approach combines chiral symmetry in the light sector with heavy-quark flavor symmetry, while Cabibbo–Kobayashi–Maskawa (CKM) matrix elements are incorporated as spurions that encode explicit symmetry breaking. Restricting to charged-current interactions, we systematically organize the leading-order current–current operators at dimension six and identify the relevant operator structures governing fully-leptonic, semi-leptonic, and hadronic decays. This framework reproduces known heavy-quark scaling relations for decay constants and form factors in agreement with expectations from heavy quark effective theory, providing nontrivial consistency checks. Operators responsible for hadronic transitions are further classified into double-trace operators and single-trace operators. These single traces, interestingly, often capture several higher order corrections, non-factorizable effects etc. We check for consistencies for both single-trace and double-trace operators demanding that the resulting amplitudes should satisfy established isospin sum rules. As an application, we analyze the decay modes $B\to K + \eta_c /\eta'/\eta $. We find that these processes receive contributions from a host of non-trivial processes such as mixing between various states, non-perturbative QCD parameters such as the heavy quark condensates, non-factorizable effects, etc, apart from the straightforward perturbative $W$ exchange diagrams in the quark picture. Our set-up neatly captures all of these effects. The phenomenological model we provide here is a symmetry-guided, hadron-level description of charged-current processes and offers a complementary perspective to conventional quark-level approaches, with a natural avenue for incorporating non-factorizable effects.
}

\maketitle
\section{Introduction}

An essential difficulty in deriving physical quantities in quantum chromodynamics (QCD) or the theory of strong interaction is the fact that one encounters non perturbative effects at long distances (see, for example \cite{Lucha:1991vn,Bali:2000gf,Alkofer:2006fu,Brambilla:2014jmp} for more details and some examples). Indeed, even electroweak transitions between hadrons are affected because of QCD, which makes comparing experimental observations with theoretical estimations in the Standard Model (SM) of particle physics challenging. While there is no alternative to lattice simulations when precision computations are concerned (see \cite{FlavourLatticeAveragingGroupFLAG:2024oxs,Aoki:2016frl,Tsang:2023nay,USQCD:2022mmc} for the current state of the lattice results relevant to flavor physics), one cannot overlook the contributions of various analytical frameworks based on (approximate) symmetries as far as developing our \emph{understanding} goes. A good example is the chiral perturbation theory or $\chi$-PT, which describes the physics of $8$ light pseudo-scalar mesons below the mass of the $\rho$-meson. In $\chi$-PT, these mesons are understood to be pseudo-Nambu-Goldstone Bosons (pNGBs) associated with spontaneous breaking of the chiral symmetry $SU(3)_L \times SU(3)_R \rightarrow SU(3)_V$. Note that quark masses and electromagnetic interactions that explicitly break the chiral symmetry leave their tell-tale signatures in the spectrum as well. Nevertheless, by exploiting these approximate symmetries $\chi$-PT describes the infrared (IR) of QCD in terms of a systematic expansion of operators with unknown low energy coefficients (LECs) has been invaluable in our \emph{understandings}. We refer the reader to \cite{Pich:1995bw,Ecker:1994gg,Scherer:2002tk} for a detailed review of the subject.

Physics is more subtle when one considers mesons containing heavy $b$ and $c$ quarks. Naively, in the physical limit (i.e., $m_b > m_c > \Lambda $, where the scales $m_{b,c}$ represent masses for $b$ and $c$ quarks respectively, and $\Lambda$ is the scale of QCD), all flavor symmetries are broken by heavy flavor masses. However, careful considerations find useful heavy quark flavor symmetries which, when coupled with emergent spin symmetries, plays the most crucial role in finding a calculable framework of heavy-light mesons (namely, heavy quark effective theory or HQET). The crucial observation stems from the fact that at $m_b, m_c \gg \Lambda$, the exact value of the heavy quark mass is largely irrelevant as far as decays to light mesons are concerned. In fact, one not only recovers the $SU(2)$ flavor of the heavy quarks, but also a further $SU(2)$ spin symmetry since the gluons decouple from the quark spin in the same limit. Ultimately, a resultant spin-flavor symmetry $SU(4)$ provides the underlying structure for the HQET. One may refer to \cite{Politzer:1988bs,Eichten:1989zv,Georgi:1990um,Grinstein:1990mj,Neubert:1993mb,Bigi:1997fj,Manohar:2000dt,Mannel:2004ce} for the construction and usage of such an effective field theory.  

Consider, for example, electroweak decays of heavy-light mesons, which provide a unique platform for studying both weak and strong interactions. The usual approach is to begin with the quark level weak Hamiltonian (a systematic expansion of effective operators as the electroweak gauge bosons $W/Z$ are  integrated out) which is renormalized down to the scale $m_b$ \cite{Buchalla:1995vs,Buras:1998raa}. The amplitude of transition is simply related to the matrix elements of these operators among the (asymptotic) hadronic states (namely, the form factors). Arguments of spin-flavor symmetries allow one to relate various matrix elements, to characterize them via Lorentz structure decomposition and spin, parity assignments, etc (see for example \cite{Isgur:1988gb,Isgur:1989vq,Isgur:1990kf,Chernyak:1983ej,Wirbel:1985ji,Bauer:1986bm,Stoler:1993yk,Falk:1990yz,Stech:1995ec,Soares:1996vs,Sterman:1997sx,Charles:1998dr}). In case of fully hadronic electroweak decays, the assumption/ansatz of factorization often brings additional simplicities and allows for complicated matrix elements to be given in terms of (simpler) factors \cite{Chau:1990ay,Blok:1992na,Neubert:1997uc,Cheng:1998uy,Ali:1998eb}. Typically, the amplitude can be written as a product of factors or quantities which can be calculated separately. The same is true for semi-leptonic processes as well, thereby enabling the amplitude to be expressed in a very simple form. This framework, coupled with data from lattice simulation, results from QCD and light cone sum rule (LCSR) (see \cite{Shifman:1978bx,Shifman:1978by,Reinders:1984sr,Narison:1989aq,Khodjamirian:2020btr,Colangelo:2000dp,Ball:2004ye,Ball:2004rg} for details and sample calculations, including subtle issues), etc, paves the way for SM estimation of rates for various electroweak transitions in the heavy-light mesons and paves the way for comparison of experimental results with SM predictions. 

There are, however, non-factorizable pieces which necessarily involve convolutions. These are rather difficult to calculate and appear as bottlenecks in theoretical estimations. In addition, there are power corrections that bring with them additional complications and theoretical uncertainties (see, for example, \cite{Jager:2012uw,Chobanova:2017ghn,Descotes-Genon:2014uoa,Jager:2014rwa,Ciuchini:2020gvn,Altmannshofer:2026cwk}). All these theoretical obstacles mar an unambiguous interpretation of the result and often obscure a direct and simple minded comparison with the experimental observations. Given such an inherent issue, one often proceeds via constructing theoretically clean observables that have significantly lower sensitivity to these hadronic uncertainties. However, there are many important quantities that are largely contaminated by these effects. Because of this, even when there are signs of deviations from theoretical expectations, one is often skeptical of interpreting these deviations as hints of possible new physics (NP) beyond the Standard Model (BSM) of particle physics. Take, for example, decays $B \to K^{(*)}\ell^+\ell^-$. Although optimized angular variables~\cite{Kruger:1999xa,Altmannshofer:2008dz,Descotes-Genon:2012isb,Matias:2012xw,Egede:2008uy,Hiller:2003js,Hiller:2014yaa,Descotes-Genon:2013vna} are considered which, by construction, are less sensitive to hadronic issues, the branching ratios themselves have high sensitivity to form factors. Intermediate charm quark loops bring in additional complications \cite{Ciuchini:1997hb,Bauer:2004tj,Ciuchini:2015qxb,Ciuchini:2021smi}. These give rise to not only non-factorizable soft-gluon effects resulting in non-trivial $q^{2}$-dependent corrections \cite{Khodjamirian:2010vf,Lyon:2014hpa,Gubernari:2020eft,Mahajan:2024xpo}, but also rescattering amplitudes (for a recent estimate on such effects, see \cite{Isidori:2024lng}) in different regions of the phase space. Attempts to estimate some of these effects involve going way beyond the usual quark-gluon level computations and having a hadron level description. 

In a similar vein, the decay modes of the $B$-meson to $\eta'/\eta$ in association with the $K$ meson present another level of difficulty. The usual procedure of estimating these modes starting with quark level operators which is reasonably successful for a plethora of other modes fails to reproduce the experimentally observed hierarchy between $B \to K \eta$ and $B \to K \eta'$. A number of ad-hoc mechanisms have been proposed ranging from the "intrinsic-charm" content of $\eta'$~\cite{Halperin:1997as,Halperin:1997ma,Shuryak:1997xd} to off-shell gluon production coupled with the QCD anomaly ~\cite{Atwood:1997bn,Hou:1997wy,Ali:1997ex,Ali:1997nh,Chen:1999nxa}  to mixing among pseudo-scalars $\eta'$-$\eta_c$~\cite{Fritzsch:1997ps,Yuan:1997ts} (see also \cite{Bagchi:1999dx,Chao:1989yp,Chao:1990im,Feldmann:1998vh} for mixing among the pseudo-scalar mesons, particularly when $\eta_c$ is included). All such effects are quite non-trivial to compute in the quark picture and often obscure the complete picture of the meson physics.    

In this work, we advocate for a complementary framework that employs intermediate level operators constructed using the meson fields themselves. We draw some of the inspiration from a "QCD-like" theory that result from a fully supersymmetric QCD (SQCD) as supersymmetry is broken via a small anomaly mediated supersymmetry breaking (AMSB). It was shown previously~\cite{Murayama:2021xfj, Csaki:2022cyg} that this theory is characterized by a QCD-like vacuum with a spontaneous break of the chiral symmetry (depending on the number of the color ($N$) and the flavor ($F$), of course). Ref.~\cite{Csaki:2025atr} even showed that the IR for $N=F=3$ is \emph{derived} to be the $\chi$-PT with \emph{predicted} LECs. While it remains to be seen whether the near-supersymmetry limit of this theory (where predictions are made) is continuously connected to the non-supersymmetric limit (i.e., the QCD)~\cite{Kondo:2025njf}, it is clear that computations on this QCD-like theory can provide qualitatively new understandings of the strong dynamics of QCD. Indeed, previous studies on AMSB+SQCD have shed new light on the physics of the pseudoscalar $\eta'$~\cite{Csaki:2023yas}, on the spectrum of scalars in QCD~\cite{Kondo:2025njf}, on the applicability of the large $N$ limit of QCD~\cite{Csaki:2025atr}, the strong CP problem of QCD~\cite{Csaki:2025atr}, etc. On a curious note, an exercise of decoupling a flavor in this theory by analytically continuing a small quark mass to a large value yields a theory of heavy-light mesons, heavy-heavy mesons, along with light pNGBs. It is characterized by non-trivial mixings among states with the same quantum numbers and, more importantly, given by a relativistic Lagrangian consisting of operators with the meson fields~\cite{Csaki:future}.     

 It is not surprising that the low energy dynamics of QCD could be described by operators with heavy-light and heavy-heavy meson field themselves. Attempts to maximize the use of chiral symmetry and heavy quark symmetries while using the pseudo-scalar fields for the heavy-light mesons as degrees of freedom (namely, the heavy hadron $\chi$-PT) exist in the literature. Although these ideas first appeared more than three decades ago (see, for example,~\cite{Wise:1992hn,Yan:1992gz,Burdman:1992gh,Cho:1992gg}) and were successful in part (see \cite{Cheng:1992xi,Grinstein:1992qt,Cho:1992cf,Casalbuoni:1996pg,Falk:1993fr,Bardeen:1993ae,Nowak:1992um,Ebert:1994tv} for a partial list of references where heavy hadron $\chi$-PT was employed), they were not pursued as actively as one had hoped. The QCD factorization methods, emerging in the early 2000s~\cite{Beneke:1999br,Beneke:2000ry,Beneke:2001ev,Keum:2000ms,Yeh:1997rq,Li:1994iu,Beneke:2001at,Bauer:2000yr,Bauer:2001yt}, provided a clean way to calculate the observables and, deservingly, became the chosen method for phenomenology. 

In this work, we make definitive progress in this direction. Using the pseudoscalar meson fields as degrees of freedom, chiral and heavy quark flavor as symmetries, elements of the Cabibbo–Kobayashi–Maskawa (CKM) matrix as spurions that explicitly break these symmetries, we categorized and tabulate leading order (of order Fermi constant or $G_F$), current-current interactions. Note that in this work, we restrict ourselves to charged-current only, which represents operators generated by a single $W$ exchange. As we show later, we not only manage to write all tree level $W$-exchange processes, but also capture several higher order processes associated with gluon exchanges, non-factorizable effects, etc. 
Since we omit current-current operators with neutral currents, we miss a class of interesting operators. In the quark picture, these neglected operators can be thought of associated with penguin diagram topologies with $W$ as one of the internal lines of the loop ( i.e. QCD and electroweak penguins, and box diagrams). The construction of these operators is rather straight-forward, and we leave it for future endeavors when we discuss neutral currents. 

In this work, we also restrict ourselves to only flavor symmetries of the heavy quarks and do not exploit the existing spin symmetry. This stops us from extending the framework to include spin one vector mesons. The main reason for narrowing down the scope of this work is the sheer number of operators that arise even at the leading order in $G_F$. Concentrating only on the pseudo-scalars, we systematically explore the space of operators at order $G_F$. As we show later in a follow up work~\cite{chakraborty:future}, the entire series of operators can be extended to make heavy flavor and spin symmetry manifest following the prescription given in~\cite{Burdman:1992gh}. 

We emphasize that the present work goes beyond the earlier works on heavy hadron $\chi$-PT theory in three essential respects: (a) the present construction allows for a straightforward and efficient way to write down amplitudes for a heavy meson decaying into light final states like $B \to K\pi$ as well as heavy-light final states like $B \to D\pi$;  (b) for the sake of completeness we find it necessary to incorporate heavy-heavy pseudo-scalars mesons in our model; and (c) the set of operators (listed below) goes way beyond the earlier works, and includes operators which directly contribute to the non-factorizable and/or sub-leading terms. At this juncture, we reiterate that even though parts of some sub-leading pieces of gluonic penguin effects are directly built in our model, we do not include operators that correspond to electroweak penguins and box diagrams as well as the leading gluonic penguins that effectively contribute to neutral current processes, since the inclusion of flavor preserving and flavor changing neutral current operators and related phenomenology will be reported elsewhere.   

In summary, we construct all operators involving pseudoscalar mesons (the pNGBs, lowest lying heavy-light, as well as heavy-heavy mesons) at leading order in $G_F$, which can be utilized to estimate charged current decays of heavy mesons. Once all operators are tabulated, we find that there are $8$ operators at the lowest order. Of these, one operator is familiar and appears in $\chi$-PT, $4$ of these give rise to fully leptonic decays of heavy-light mesons, and one operator yields the fully leptonic decay of $B_c$. Apart from these, these operators generate a host of semi-leptonic decays. These decays provide ample opportunities to check for consistencies using known results from heavy quark spin symmetry. We utilized known results from HQET and reproduced a number of  scalings and relations among these transitions. The physics of fully hadronic electroweak decays of mesons is, on the other hand, rather complicated. We find $28$ double trace operators and $40$ single trace operators that describe these decays at the leading order in $G_F$. We employ isospin sum rules as a measure of consistency checks. Finally, we analyze decays $B\to K + \eta_c/\eta'/\eta$ within our framework, which are impacted non-trivially by mixings in the $\eta_c/\eta'/\eta$-system, non-perturbative QCD parameter such as the heavy quark condensates, non-factorizable effects, apart from the straightforward perturbative $W$ exchange diagrams in the quark picture. We demonstrate that all these effects neatly show up in our set-up can in a tractable manner. In fact, the sizes of each of these effects can be estimated if one makes an additional ansatz of naturalness, which suggests that there are no large cancellations that ultimately yield the hierarchies observed in these three modes of decays.    

Before we proceed, note that effective field theories (EFTs) of this kind, strictly speaking, are suitable for describing the corners of the phase space where the pNGBs carry low momentum (i.e., soft). In modes like $B \to P \ell\nu(\ell)$, such a construction will provide a good description when the lepton pair carries a large fraction of the total energy. In case of few body fully hadronic decays, where the resultant momentum of pNGBs are relatively large, the usage of the EFT may appear problematic. This issue is well appreciated and has been raised in literature (see for examples \cite{Yan:1992gz, Burdman:1992gh}). Ref.\cite{Yan:1992gz}, in fact, argues that it is known to the practitioners in the subject that, when treated as a pole model, the chiral Lagrangian works well beyond the soft pion limit. In this work one can instead take the approach of Ref.~\cite{Burdman:1992gh}. We view the proposed set of operators as a phenomenological model of charged-current processes. Note, one way to arrive at the model (i.e., the collection of operators) is via working at a point in the parameter space where heavy quark masses are nearly degenerate and are not too heavy with respect to the QCD scale, while at the same time the light quark masses approach the QCD scale from below. As one moves towards physical masses, we expect the operators to remain the same, even though corresponding coefficients may change continuously. In our approach, we use symmetries only to identify and categorize operators and propose to use experimental-data/lattice-data/known-results from HQET, LCSR, etc to determine the coefficients in any case. 

Rest of this work is organized as follows: in Sec. \ref{sec:model} we give details of our set-up and tabulate the collection of operators we employ for describing charged current interactions of pseudo-scalar mesons; in Sec. \ref{sec:consistency_leptonic} we describe fully leptonic and semileptonic decay modes of heavy hadrons in our model after summarizing various known results/scalings from heavy quark spin symmetry considerations; in Sec. \ref{sec:consistency_hadronic} we check for consistency associated with predictions of hadronic decay modes from known isospin considerations; in Sec. \ref{sec:B2KY} we perform a phenomenological analysis of $B\to K + \eta_c/\eta'/\eta$ within our model; and finally in Sec. \ref{sec:conclusion} we conclude.

\section{A phenomenological model of pseudoscalar mesons}
\label{sec:model}

The purpose of this section is to completely specify the degrees of freedom and the symmetry structure, which allow us to build a phenomenological model of charged weak decays of pseudo-scalar hadrons. The details of our arguments behind this construction are given in \cref{sec:app1}. 

The starting ansatz is that the physics of charged electroweak decays of heavy mesons can be largely understood within the symmetry structure of 
\begin{equation}
    G_{\text{approx}} \ \equiv \ U(3)_L \times U(3)_R \times SU(2)_H \; ,
\end{equation}
broken by electroweak charges in a tractable manner. The subset $U(3)_L \times 
U(3)_R$ is the usual flavor symmetry of light quarks in the massless light flavor limit. The $SU(2)_H$ piece is the vectorial heavy flavor symmetry of heavy quarks (charm and bottom), which can be understood either in the degenerate bottom and charm mass limit (i.e., $m_c \to m_b$) or in the heavy quark limit (i.e., $m_c, m_b \to \infty $). In this section, we use the notation $L, R$, and $V_H$ to denote elements of $U(3)_L, U(3)_R $, and $SU(2)_H$, respectively. 

It is not so surprising to talk about flavor symmetric charged-weak decays even though the symmetry is explicitly broken in charged-current interactions. Take, for example, the quark Lagrangian and consider interactions among the $W$-boson, the left-handed $c$-quark, and the three left-handed light flavors (namely, $u,d$ and $s$ quarks). The imposition of $SU(3)_L$ flavor symmetry demands that the $c$-quark couples to all light flavors equally. In reality, there are no charged-current interactions between the $c$-quark and the $u$-quark (because of electromagnetic charge considerations), and even the couplings of the $c$-quark with $d$ and $s$ differ from each other because of CKM matrix-elements. We overcome this difficulty by introducing (electroweak) charge matrices (containing various CKM matrix-elements arranged accordingly), which are treated as "spurions" (i.e., are assigned spurious transformation properties under flavor symmetries) to make the charge-current interactions flavor symmetric. The details of this procedure are given in \cref{sec:app1}.  

Let us reiterate, the electroweak part of the quark Lagrangian can accommodate $G_{\text{approx}}$ when electroweak charges are treated as spurions. Following \cref{sec:app1}, we need to consider four (electroweak) charge matrices, namely, 
\begin{equation}
    \hat{Q}_{qq} \equiv 
        \begin{pmatrix} 0 & 0 & 0 \\ 
                        V_{\bar{d}u} & 0 & 0 \\    
                        V_{\bar{s}u} & 0 & 0  
        \end{pmatrix}\, ,  \quad
    \hat{Q}_{QQ} \equiv 
        \begin{pmatrix} 0 & 0  \\ 
                        V_{\bar{b}c} & 0   
        \end{pmatrix}\, ,  \quad         
    \hat{Q}_{qQ} \equiv 
        \begin{pmatrix} 0 & 0  \\ 
                        V_{\bar{d}c} & 0  \\    
                        V_{\bar{s}c} & 0   
        \end{pmatrix}\, ,  \quad  
    \hat{Q}_{Qq} \equiv 
        \begin{pmatrix} 0 & 0 & 0 \\ 
                        V_{\bar{b}u} & 0 & 0 
        \end{pmatrix}\, , 
\label{eq:defWcharges}
\end{equation}
and their spurious transformation under $G_{\text{approx}}$ given by 
\begin{equation}
    \hat{Q}_{qq}  \rightarrow L \hat{Q}_{qq} L^\dag \; ,\quad 
    \hat{Q}_{QQ}  \rightarrow V_H \hat{Q}_{QQ} V_H^\dag \; ,\quad
    \hat{Q}_{qQ}  \rightarrow L \hat{Q}_{qQ} V_H^\dag \; , \quad
    \hat{Q}_{Qq}  \rightarrow V_H \hat{Q}_{Qq} L^\dag \;.
 \label{eq:defWchargestransformation}   
\end{equation}
The key insight for this work is that one can use these charge matrices and their spurious transformations even in a meson picture to construct a phenomenological Lagrangian useful for describing the weak decays of heavy mesons. 

Apart from the charge matrices discussed above, the basic ingredients in our set-up are color-neutral pseudo-scalar fields that create/annihilate corresponding hadrons. We utilize three matrix fields : $\Sigma$ containing pions in the non-linear representation, the field $H$, which contains heavy-light bound states, and finally the field $\Sigma_H$, which contains the heavy-heavy bound states. 

We begin with the nine light pseudoscalar mesons (pNGBs). We use the same notation as in the $\chi$-PT. 
\begin{equation}
\begin{split}
 \Sigma \ \equiv \ \frac{1}{2} \: f_\pi U_\pi \; , \quad \text{where} \quad  
        U_\pi \equiv \exp{\left( \frac{2\iu\pi^a t^a}{f_\pi } \right)} \quad
        \text{and} \quad \Sigma \ \xrightarrow[]{G_{\text{approx}}} \ L \Sigma R^\dag\,. \\
  \text{In particular,} \qquad 
      \pi^a t^a = 
    \begin{pmatrix}
        \frac{\pi^0}{\sqrt{2}} + \frac{\eta_8}{\sqrt{6}} & \pi^+ & K^+ \\
        \pi^- & -\frac{\pi^0}{\sqrt{2}} +\frac{\eta_8}{\sqrt{6}} & K^0 \\
        K^- & \bar{K}^0 & -\frac{2\eta_8}{\sqrt{6}}
    \end{pmatrix}\; + \frac{1}{\sqrt{3}} \eta_0 \mathbb{1}\,.
\end{split}
\label{eq:defCompFieldsSigma}        
\end{equation}
Note that in \eqref{eq:defCompFieldsSigma} we use the normalization $\text{Tr}\left( t^a t^b\right) = \delta^{ab}$ for the generators, where $a$ runs over $0, \ldots, 8$. At $\Order{p^2}$, all fields listed in \cref{eq:defCompFieldsSigma} receive masses proportional to quark masses. The same operator also generates a mixing (from a mixed-mass operator) between $\eta_0$ and $\eta_8$. However, note that the dynamics of non-perturbative QCD additionally breaks the axial $U(1)$ and generates a separate potential (i.e., a mass of order $\Lambda$) for the $\eta_0$ field. Since the mixing is $\Order{m_q}$ suppressed, the heavier mass eigenstate (the $\eta'$ field) resides mostly in $\eta_0$, while the lighter $\eta$ is predominantly given in $\eta_8$\footnote{Here we neglect mixing with $\pi^0$, since it is additionally suppressed by a factor $m_d/m_s$.}. Later in \cref{sec:B2KY} we expand this understanding further.

Next we move to the lowest lying heavy-heavy pseudo scalar fields. As mentioned above, these are contained in the matrix $\Sigma_H$, which, as expected, is in the adjoint + singlet representation of the heavy flavor group. 
\begin{equation}
    \Sigma_H \ \equiv \ \begin{pmatrix}
        \Lambda_c + \iu \eta_c & \iu B_c^{+} \\
        \iu B_c^{-} & \Lambda_b + \iu \eta_b \\
    \end{pmatrix} \qquad \text{with} \qquad 
    \Sigma_H \ \xrightarrow[]{G_{\text{approx}}} \ V_{H} \Sigma_{H} V_{H}^\dag  \;.
\label{eq:defCompFieldsSigmaH}     
\end{equation}
In \eqref{eq:defCompFieldsSigmaH}, $\Lambda_c$ and $\Lambda_b$ are real parameters. As we argue later, these parameters are representative (proportional to) the condensates $\langle \bar{c} c \rangle$ and $\langle \bar{b} b \rangle$, respectively. Although we will leave the value of these parameters to be determined phenomenologically, we do expect each of these $\Lambda$ parameters to scale to zero as the corresponding quark masses $m_c$ and $m_b$ are taken to infinity. In the limit $m_b \to m_c$, we expect $\Lambda_b \to \Lambda_c$ and $SU(2)_H$ to become exact, with all fields in $\Sigma_H$ degenerate. As the theory moves away from the $SU(2)_H$ symmetric point, one expects $\Lambda_c$ to pull away from $\Lambda_b$ with the mass gap between different fields in $\Sigma_H$ increasing.      

Representing the lowest-lying heavy-light pseudo-scalar states within our set-up is rather non-trivial as they do not transform in a nice fashion as the others. To see this, consider the operator $\bar{q} \gamma_5 Q$ (containing all the quantum numbers of a heavy-light pseudo-scalar field), where $q$ and $Q$ represent the light and heavy quarks, respectively. Under $G_{\text{approx}}$, different components of $\bar{q} \gamma_5 Q$ transform differently. To be explicit, $\bar{q}_L \gamma_5 Q \to V_H \left( \bar{q}_L \gamma_5 Q \right) L^\dag$, whereas  $\bar{q}_R \gamma_5 Q \to V_H \left( \bar{q}_R \gamma_5 Q \right) R^\dag$, i.e., the operator explicitly breaks $G_{\text{approx}}$. 

Note that the purpose of relying on $G_{\text{approx}}$ in this work is only to identify and categorize \emph{all} operators relevant for charged-weak decays at the leading order. For this reason, we introduce a neat trick. We begin with two seemingly unrelated fields in the given representations under $G_{\text{approx}}$, namely $\Phi \to L \Phi V_H^\dag $ and $\Phi_c \to V_H \Phi R^\dag $. We construct the complete set of operators using $\Phi$ and $\Phi_c$, which are finally substituted by the matrix field $H$ with the rules $\Phi_c \to H$ and $\Phi \to H^\dagger$. The components of $H$ are given as: 
\begin{equation}
    H \equiv \begin{pmatrix}
        D & D^{+} & D^{+}_s \\
        B^{-} & \bar{B^0} & \bar{B}_s \\
    \end{pmatrix}\,. 
\label{eq:defCompFieldsH}         
\end{equation}
Noting that $H$ has homogeneous transformation properties under the fully vectorial $SU(3)_V \times SU(2)_H$, where $SU(3)_V$ represents the diagonal subgroup of the light flavor $SU(3)_L\times SU(3)_R$ one could have chosen it instead of $G_{\text{approx}}$. In this work, we prefer to work with the full $G_{\text{approx}}$ because we find it to be more illuminating.  

We now have enough information to discuss operators responsible for charged weak decays of pseudoscalar mesons. Our ansatz is that each of these component fields is characterized by small anomalous dimensions around the engineering dimensions one (i.e., these are necessarily weakly coupled) at energies where we deploy this set-up. This allows us to construct these operators systematically from dimensional considerations. As described earlier in this section, all our operators are product of two charged-currents, where each current is characterized by the mass dimension $3$ in the leading order. Therefore, not surprisingly, we find that the leading operator arises at $D=6$. All such operators must have a single factor of the Fermi factor $G_F$, which fixes its mass dimension. 

Consider first charged current-current operators with one of the currents consisting of leptons. The remaining current, therefore, must be invariant under symmetry transformations of the global $G_{\text{approx}}$, i.e., invariant products of a single factor of the charge matrices in \cref{eq:defWcharges}, two fields and one momentum. We find eight such different operators, which are listed in \cref{table:hadroniccurrents}. The relevant part of the Lagrangian describing the leptonic decays of mesons, therefore, can be given as   
\begin{equation}
\begin{gathered}
    \mathcal{L}_{\text{CC}} \ \supset \ \frac{4G_F}{\sqrt{2}} \:
           \sum_{i=1}^8 \:  \left( \mathcal{K}_i \: \Oph_i \: \Opl \ + \ \text{h.c.} \right) \;  \ = \ 
            \frac{4G_F}{\sqrt{2}} \ \Oph  \: \mathcal{K}^T \:  \Opl \ + \ \text{h.c.} \\
    \text{where} \quad 
    \Opl \ = \ \sum_{\ell} \left( \nu^\dag_{\ell} \bar{\sigma}^\mu \ell \right) \; ,  \\
    \Oph \ = \ \begin{pmatrix}
                    \Oph_1, \Oph_2, \ldots , \Oph_8 
            \end{pmatrix}   \; ,
    \quad \text{and} \quad
    \mathcal{K} \ = \ \begin{pmatrix} 
                  \mathcal{K}_1, \mathcal{K}_2, \ldots, \mathcal{K}_8      
                \end{pmatrix} \; , 
\end{gathered}            
    \label{eq:operatorssemileptonic}       
\end{equation}
where, we take $\Opl$ to be the leptonic current (summed over flavor), $\Oph$ to be a vector of eight operators listed in \cref{table:hadroniccurrents}, and $\mathcal{K}$ to be the corresponding vector of eight coupling constants $\mathcal{K}_i$. We employ the notation $\tr{\cdots}$ to denote a trace over flavor indices. Note again that in our convention the charge matrices become parts of $\Oph_i$ which makes each of these operators to be $G_{\text{approx}}$ invariant.   
\begin{table}[ht]
    \centering
    \begin{tabular}{|c|c||c|c|}
    \hline
         $\Oph_1$ &  $\tr{ \Sigma^\dag \: \hat{Q}_{qq}  \: 
            \overleftrightarrow{\partial}_\mu \Sigma } $ &
             $\Oph_2$ &  $\tr{ \Sigma_H^\dag \: \hat{Q}_{QQ} \:  
                \overleftrightarrow{\partial}_\mu \Sigma_H} $ \\
         $\Oph_3$ &  $\tr{ H^\dagger \: \hat{Q}_{QQ} \: 
                \overleftrightarrow{\partial}_\mu H } $   &
         $\Oph_4$ &  $\tr{ H \: \hat{Q}_{qq} \: 
                \overleftrightarrow{\partial}_\mu H^\dagger} $  \\   
         $\Oph_5$ &  $\tr{ \Sigma^\dag \: \hat{Q}_{qQ} \:
                 \overleftrightarrow{\partial}_\mu H} $ &
         $\Oph_6$ &  $\tr{ \Sigma_H^\dag \: \hat{Q}_{Qq} \:
                 \overleftrightarrow{\partial}_\mu H^\dagger } $ \\
         $\Oph_7$ &  $\tr{ H^\dagger \: \hat{Q}_{Qq} \:
                 \overleftrightarrow{\partial}_\mu \Sigma } $ & 
         $\Oph_{8}$ &  $\tr{ H \: \hat{Q}_{qQ} \:
                 \overleftrightarrow{\partial}_\mu \Sigma_H } $ \\ 
                 \hline     
    \end{tabular}
    \caption{The list of operators specified in \cref{eq:operatorssemileptonic}.}
    \label{table:hadroniccurrents}
\end{table}

Not surprisingly, the same operators in \cref{table:hadroniccurrents} can be used to construct all-hadronic current-current interactions (or at least a part of it), since these operators represent symmetry-invariant hadronic currents at the lowest mass dimension. The product of one operator from the list with the Hermitian conjugate of another one satisfies all our criteria. This procedure gives rise to a set of distinct operators in the Lagrangian, built out of the vector $\Oph$ contains $8$ unique operators. Since each current is $G_{\text{approx}}$ invariant, these operators are characterized by two independent traces over $G_{\text{approx}}$ indices. From now on, we will refer to these as \emph{double-trace operators}. 

There are also \emph{single-trace operators} in the same order. The way to construct these systematically is via the charge matrices. Notice that at the lowest order, the underlying physics we are attempting to model is due to a single $W$-propagator (and hence a single $G_F$ factor), which suggests the presence of two charge matrices (in fact, one of the $\hat{Q}$ matrices and one $\hat{Q}^\dagger$), each accompanied by a double-sided derivative and surrounded by fields from both sides. An algorithmic way to find all such operators starts with placing any of the combinations of charge matrices, and then one figures out the field content using the spurious transformation properties of fields and charge matrices. \cref{table:singletrace} lists all such operators.  

Combining both double and single traces, we, therefore, identify the Lagrangian at the leading order that gives rise to hadronic decay of pseudo-scalar mesons to be  
\begin{equation}
\begin{gathered}
    \mathcal{L}_{\text{CC}}  \ \supset \  \frac{4G_F}{\sqrt{2}} \Bigg\{ \sum_{ i,j = 1}^8 
            \mathcal{C}^i_j \: \Oph_i  \: {\Oph_j}^\dag    \ + \ 
            \sum_{a=1}^{14} \mathcal{E}_a \:  \Ophh_a  \ + \ 
            \sum_{a=15}^{40} \left( \mathcal{E}_a \:  \Ophh_a \ + \ \text{h.c.} \right)
                    \Bigg\}\;,  \\ 
            \ = \  \frac{4G_F}{\sqrt{2}} \Bigg\{  \Oph \: \mathcal{C} \: {\Oph}^\dag \ + \ 
                        \Ophh_R \: \mathcal{E}_R^T \ + \  
               \left(  \Ophh_C \: \mathcal{E}_C^T  \ + \ \text{h.c.} \right) 
                       \Bigg\} \; ,  \\
        \text{where} \quad  
        \Ophh_R \ = \ \begin{pmatrix}
                    \Ophh_1,  \ldots , \Oph_{14} 
            \end{pmatrix} \, ,  \quad
        \Ophh_C \ = \ \begin{pmatrix}
                    \Ophh_{15},  \ldots , \Oph_{40} 
            \end{pmatrix} \, , \\
        \mathcal{E}_R \ = \ \begin{pmatrix} 
                  \mathcal{E}_1, \ldots, \mathcal{E}_{14}
                \end{pmatrix} \, ,  \quad
        \mathcal{E}_C \ = \ \begin{pmatrix} 
                  \mathcal{E}_{15}, \ldots, \mathcal{E}_{40}
                \end{pmatrix} 
                \; ,  \quad \text{and finally} \quad 
                    \left( \mathcal{C}^i_j \right)^* = \mathcal{C}^j_i \; . 
    \label{eq:operatorsfullyhadronic}
\end{gathered}    
\end{equation}
In the above, $\mathcal{C}^j_i$ and $\mathcal{E}_a$ are Wilson coefficients. The operators $\Ophh_{1, \ldots , 14}$ are real and so are the coefficients $\mathcal{E}_{1, \ldots , 14}$. The rest of the single trace operators are complex. In the second and third lines of \cref{eq:operatorsfullyhadronic} we use a more compact notation to rewrite the operators. By definition, the matrix $\mathcal{C}$ is Hermitian and $\mathcal{E}_R$ is a row matrix consisting of real elements. 

\begin{table}
    \centering
    \begin{tabular}{|c|c||c|c|}
    \hline 
        $\Ophh_1$ & $\tr{\Sigma_H^\dag \hat{Q}_{QQ} \darrowD  \Sigma_H \ 
            \Sigma_H^\dag  \hat{Q}_{QQ}^\dag  \darrowU  \Sigma_{H} }$  &
        $\Ophh_{2}$ &   $\tr{H^\dagger  \hat{Q}_{QQ} \darrowD  H \ 
            H^\dagger   \hat{Q}_{QQ}^\dag  \darrowU  H }$ \\
        $\Ophh_{3}$ & $\tr{H \hat{Q}_{qQ} \darrowD  \Sigma_H \ 
            \Sigma_H^\dag  \hat{Q}_{qQ}^\dag  \darrowU  H^\dagger }$  & 
       $\Ophh_{4}$ &   $\tr{\Sigma^\dag \hat{Q}_{qQ} \darrowD  \Sigma_H \ 
            \Sigma_H^\dag  \hat{Q}_{qQ}^\dag  \darrowU  \Sigma }$ \\
       $\Ophh_{5}$ & $\tr{H \hat{Q}_{qQ} \darrowD  H \ 
            H^\dagger  \hat{Q}_{qQ}^\dag  \darrowU  H^\dagger }$  & 
       $\Ophh_{6}$ &     $\tr{\Sigma^\dag \hat{Q}_{qQ} \darrowD  H \ 
            H^\dagger  \hat{Q}_{qQ}^\dag  \darrowU  \Sigma }$ \\     
       $\Ophh_{7}$ & $\tr{\Sigma_H^\dag \hat{Q}_{Qq} \darrowD  \Sigma \ 
            \Sigma^\dag  \hat{Q}_{Qq}^\dag  \darrowU  \Sigma_{H} }$  & 
       $\Ophh_{8}$ &     $\tr{H^\dagger \hat{Q}_{Qq} \darrowD  \Sigma \ 
            \Sigma^\dag   \hat{Q}_{Qq}^\dag  \darrowU  H }$ \\
       $\Ophh_{9}$ & $\tr{\Sigma_H^\dag \hat{Q}_{Qq} \darrowD  H^\dagger \ 
            H  \hat{Q}_{Qq}^\dag  \darrowU  \Sigma_{H} }$ & 
       $\Ophh_{10}$ &     $\tr{H^\dagger \hat{Q}_{Qq} \darrowD  H^\dagger \ 
            H  \hat{Q}_{Qq}^\dag  \darrowU  H }$\\     
       $\Ophh_{11}$ & $\tr{ H \hat{Q}_{qq} \darrowD  \Sigma \ 
            \Sigma^\dag  \hat{Q}_{qq}^\dag  \darrowU  H^\dagger  }$  & 
       $\Ophh_{12}$ &       $\tr{\Sigma^\dag \hat{Q}_{qq} \darrowD  \Sigma \ 
            \Sigma^\dag   \hat{Q}_{qq}^\dag  \darrowU  \Sigma }$ \\
       $\Ophh_{13}$ &   $\tr{ H \hat{Q}_{qq} \darrowD  H^\dagger \ 
            H  \hat{Q}_{qq}^\dag  \darrowU  H^\dagger }$ & 
       $\Ophh_{14}$ &       $\tr{\Sigma^\dag \hat{Q}_{qq} \darrowD  H^\dagger \ 
            H  \hat{Q}_{qq}^\dag  \darrowU  \Sigma }$\\                 
    \hline \hline

        $\Ophh_{15}$ & $\tr{H^\dagger \hat{Q}_{QQ} \darrowD  \Sigma_H \ 
                H  \hat{Q}_{qq}^\dag  \darrowU  \Sigma }$ & $\Ophh_{16}$ & $\tr{\Sigma_H^\dag \hat{Q}_{QQ} \darrowD  \Sigma_H \ 
            H  \hat{Q}_{qq}^\dag  \darrowU  H^\dagger }$  \\
         
        $\Ophh_{17}$ &  $\tr{H^\dagger \hat{Q}_{QQ} \darrowD  \Sigma_H \ 
                H  \hat{Q}_{Qq}^\dag  \darrowU  H }$ & $\Ophh_{18}$ & $\tr{\Sigma_H^\dag \hat{Q}_{QQ} \darrowD  \Sigma_H \ 
            H  \hat{Q}_{Qq}^\dag  \darrowU  \Sigma_{H} }$ \\

        $\Ophh_{19}$ &   $\tr{H^\dagger \hat{Q}_{QQ} \darrowD  \Sigma_H \ 
            \Sigma_H^\dag  \hat{Q}_{qQ}^\dag  \darrowU  \Sigma }$ & $\Ophh_{20}$ & $\tr{\Sigma_H^\dag \hat{Q}_{QQ} \darrowD  \Sigma_H \ 
            \Sigma_H^\dag  \hat{Q}_{qQ}^\dag  \darrowU  H^\dagger }$   \\
         
        $\Ophh_{21}$ &    $\tr{H^\dagger \hat{Q}_{QQ} \darrowD  \Sigma_H \ 
            \Sigma_H^\dag  \hat{Q}_{QQ}^\dag  \darrowU  H }$ & 
            
        $\Ophh_{22}$ &$\tr{\Sigma_H^\dag \hat{Q}_{QQ} \darrowD  H \ 
            \Sigma^\dag \hat{Q}_{qq}^\dag  \darrowU  H^\dagger }$ \\
        $\Ophh_{23}$ &    $\tr{H^\dagger \hat{Q}_{QQ} \darrowD  H \ 
            \Sigma^\dag \hat{Q}_{qq}^\dag  \darrowU  \Sigma }$ &
        $\Ophh_{24}$ & $\tr{\Sigma_H^\dag \hat{Q}_{QQ} \darrowD  H \ 
            \Sigma^\dag \hat{Q}_{Qq}^\dag  \darrowU  \Sigma_{H} }$ \\ 
        $\Ophh_{25}$ &  $\tr{H^\dagger \hat{Q}_{QQ} \darrowD  H \ 
            \Sigma^\dag \hat{Q}_{Qq}^\dag  \darrowU  H }$ &
            
         $\Ophh_{26}$ & $\tr{\Sigma_H^\dag \hat{Q}_{QQ} \darrowD  H \ 
            H^\dagger   \hat{Q}_{qQ}^\dag  \darrowU  H^\dagger }$  \\ 
         $\Ophh_{27}$ &   $\tr{H^\dagger  \hat{Q}_{QQ} \darrowD  H \ 
            H^\dagger   \hat{Q}_{qQ}^\dag  \darrowU  \Sigma }$ &
         $\Ophh_{28}$ & $\tr{\Sigma_H^\dag \hat{Q}_{QQ} \darrowD  H \ 
            H^\dagger   \hat{Q}_{QQ}^\dag  \darrowU  \Sigma_{H} }$  \\ 

%& \\
            \hline 
%& \\            

       $\Ophh_{29}$ & $\tr{H \hat{Q}_{qQ} \darrowD  \Sigma_H \ 
            H  \hat{Q}_{Qq}^\dag  \darrowU  \Sigma_H }$  & 
       $\Ophh_{30}$ &     $\tr{\Sigma^\dag \hat{Q}_{qQ} \darrowD  \Sigma_H \ 
            H  \hat{Q}_{Qq}^\dag  \darrowU  H }$ \\
       $\Ophh_{31}$ & $\tr{H \hat{Q}_{qQ} \darrowD  H \ 
            \Sigma^\dag  \hat{Q}_{Qq}^\dag  \darrowU  \Sigma_H }$  & 
       $\Ophh_{32}$ &     $\tr{\Sigma^\dag \hat{Q}_{qQ} \darrowD  H \ 
            \Sigma^\dag  \hat{Q}_{Qq}^\dag  \darrowU  H }$ \\    
       $\Ophh_{33}$ & $\tr{H \hat{Q}_{qQ} \darrowD  \Sigma_H \ 
            H  \hat{Q}_{qq}^\dag  \darrowU  H^\dagger }$  & 
       $\Ophh_{34}$ &     $\tr{\Sigma^\dag \hat{Q}_{qQ} \darrowD  \Sigma_H \ 
            H  \hat{Q}_{qq}^\dag  \darrowU  \Sigma }$ \\
       $\Ophh_{35}$ &  $\tr{H \hat{Q}_{qQ} \darrowD  H \ 
            \Sigma^\dag  \hat{Q}_{qq}^\dag  \darrowU  H^\dagger }$  & 
       $\Ophh_{36}$ &     $\tr{\Sigma^\dag \hat{Q}_{qQ} \darrowD  H \ 
            \Sigma^\dag  \hat{Q}_{qq}^\dag  \darrowU \Sigma }$ \\        
%         & \\
         \hline 
%         & \\

       $\Ophh_{37}$ &$\tr{\Sigma_H^\dag \hat{Q}_{Qq} \darrowD  \Sigma \ 
            \Sigma^\dag  \hat{Q}_{qq}^\dag  \darrowU  H^\dagger  }$  & 
       $\Ophh_{38}$ &     $\tr{H^\dagger \hat{Q}_{Qq} \darrowD  \Sigma \ 
            \Sigma^\dag   \hat{Q}_{qq}^\dag  \darrowU  \Sigma }$ \\
       $\Ophh_{39}$ & $\tr{\Sigma_H^\dag \hat{Q}_{Qq} \darrowD  H^\dagger \ 
            H  \hat{Q}_{qq}^\dag  \darrowU  H^\dagger }$ & 
       $\Ophh_{40}$ &     $\tr{H^\dagger \hat{Q}_{Qq} \darrowD  H^\dagger \ 
            H  \hat{Q}_{qq}^\dag  \darrowU  \Sigma }$\\         
%         & \\
       \hline

    \end{tabular}
    \caption{Single Trace Operators.}
    \label{table:singletrace}
\end{table}

%-------------------------------------------

\section{Consistency checks and phenomenology of leptonic \& semi-leptonic decays} 
\label{sec:consistency_leptonic}

We devote this section to discuss the phenomenology of leptonic and semi-leptonic decays of mesons in \cref{eq:operatorssemileptonic} with operators in \cref{table:hadroniccurrents}. 

First of all, note that the first term in \cref{eq:operatorssemileptonic} is proportional to $\Oph_1 \times \Opl$, gives (semi-)leptonic decays of light mesons, and is identical to what one gets in the $\chi$-PT at leading order. Therefore, it does not require any additional discussion. The operators $\Oph_{3,4} \times \Opl$ generate semi-leptonic $B \to D$ transitions; operators $\Oph_{5,6,7,8} \times \Opl$ give leptonic decays of $B$ and $D$ mesons; operators $\Oph_{5,7} \times \Opl$ generate semi-leptonic decays of $B$ and $D$ mesons with light mesons in the final state; operator $\Oph_{2}\times \Opl$ give leptonic decay of $B_c$; and finally, operators $\Oph_{2,6,8}\times \Opl$ also generate all semi-leptonic decays of $B_c$. 

However, before we begin with phenomenological explorations of the model in \cref{subsec:pheno_leptonic_model}, we briefly discuss some of our current analytical understanding of these decays. To be specific, we review known HQET scalings of various $B/D$ form factors and decay constants in \cref{subsec:hqet_heavylight}. We also review various scalings in the context of $B_c$ decays in \cref{subsec:hqet_heavyheavy}, which are understood with the tool of heavy quark spin symmetry and using models of wave functions for initial and final mesons.      
%-----------------
\subsection{Leptonic and semileptonic decays of $B/D$ meson in HQET}
\label{subsec:hqet_heavylight}

We begin with the decay constants of $B/D$-pseudoscalar mesons (namely, $f_B$ and $f_D$, respectively), which are defined in the corresponding purely leptonic decays. In the quark picture of the SM, these decays occur entirely via Fermi operators $ \left(\bar{b}_L \, \partial_{\mu}   u_L \right) J^\mu_L$ and $ \left(\bar{d}_L \, \partial_{\mu}   c_L \right) J^\mu_L$ respectively, which allows one to clearly define the decay constants as matrix elements of these specific hadronic currents. To be specific, 
\begin{equation}
    f_B  \equiv  i \: \BState{0} \bar{u} \, \slashed{p}_B \, 
                \gamma^5 b(0) \State{B^{+}\left(p_B \right)} /m_B^2  \quad \text{and} \quad
    f_D  \equiv  i \: \BState{0} \bar{d} \, \slashed{p}_D \, \gamma^5 c(0)\State{D^{+}(p_B)} /m_D^2         \; .   
    \label{eq:defDC}
\end{equation}
One can use arguments from HQET to determine how these decay constants (of heavy-light mesons) scale with the masses of heavy quarks. The trick is to go to the non-relativistic limit, where one switches the heavy quark field by its velocity eigenstates and then expands the current in terms of expansions in non-relativistic meson fields which creates and destroys non-relativistic one particular heavy-meson states. After matching and carefully tracking the normalization of non-relativistic states, one finds the following nice results 
\begin{equation}
\begin{split}
   f_{P_Q} \times \sqrt{m_{P_Q}} \sim \text{const.} + \mathcal{O}(\alpha_s \ln m_P)\,,  \\
         \implies \qquad \frac{f_B}{f_D} \ = \ \sqrt{\frac{m_D}{m_B}} \left( 1 + \frac{\alpha_s}{\pi} 
         \ln \left( \frac{m_B}{m_D}\right) + \ldots \right) \; ,
\end{split}    
\label{eq:scaleDC}
\end{equation}
where $P_Q$ denotes the heavy-light pseudo-scalar meson containing the heavy quark $Q$ and $\ldots$ represent even higher-order terms. 

We can find similar scalings in transition matrix elements (or, form factors) in semi-leptonic decays of heavy-light mesons as well. Consider, for example, decays $\bar{B}\rightarrow \pi \ell \bar{\nu}$ and $D \rightarrow \pi \bar{e}\nu$. Using the notation in \cref{eq:scaleDC}, we define form factors as 
\begin{equation}
    \BState{\pi\left( p_\pi \right)} \bar{q} \gamma_\mu (1-\gamma_5) Q  \State{P_{Q}\left(p_{_{P_Q}} \right)} \ = \ f_{+}^{P_Q \rightarrow \pi} \left( p_{_{P_Q}} + p_\pi\right)_\mu + 
        f_{-}^{P_Q \rightarrow \pi} \left( p_{_{P_Q}} - p_\pi\right)_\mu   \; .
    \label{eq:defFF1}
\end{equation}
Once again, one can go to the HQET limit and use the velocity eigenstates of $P_Q$ (in the regions of phase space where large momentum transfer to the leptonic sector takes place, i.e., $v\cdot p_\pi \ll m_{P_Q}$). Extracting the leading $m_{P_Q}$ dependence from both sides of \cref{eq:defFF1}, one gets (at order $m_{P_Q}$)
\begin{equation}
    f_{+}^{P_Q \rightarrow \pi} + f_{-}^{P_Q \rightarrow \pi} \sim \mathcal{O}\left( \frac{1}{\sqrt{m_{P_Q}}}\right) 
        \qquad \text{and} \qquad 
    f_{+}^{P_Q \rightarrow \pi} - f_{-}^{P_Q \rightarrow \pi} \sim \mathcal{O}\left( \sqrt{m_{P_Q}}\right) \; . 
    \label{eq:scaleFF}
\end{equation}
As shown in~\cite{Isgur:1989vq, Isgur:1990yhj}, one can utilize heavy quark spin-flavor symmetry  arguments to determine the leading operators involving heavy-light mesons and light mesons. In particular, this construction necessitated the need for superfield-representations that combine the pseudo-scalar and the (nearly degenerate) vector meson degrees of freedom into a single $4\times 4$ matrix-field in the spinor-space for unifying heavy flavor-spin symmetry. However,  the leading effective operator involving $P_Q, \pi$ and the leptonic current does not generate $\left( f_{+} - f_{-} \right)$, which arises via the leading coupling between $P_Q, P^{*}_Q$ and $\pi$ (often denoted in the literature by $g_\pi$). In particular, one finds, at the leading order in $m_{P_Q}$:
\begin{equation}
\begin{split}
        f_{+}^{P_Q \rightarrow \pi} + f_{-}^{P_Q \rightarrow \pi}  \ &= \ \frac{f_P}{f_\pi}
        \left( 1 -  \frac{g_\pi \times v\cdot p_\pi }{v\cdot p_\pi + m_{P^*} - m_P} \right)\;, \\ 
        f_{+}^{P_Q \rightarrow \pi} - f_{-}^{P_Q \rightarrow \pi}  \ &= \ \frac{f_P}{f_\pi}
         \frac{ g_\pi\times  m_P }{v\cdot p_\pi + m_{P^*} - m_P} \;.
\end{split}        
\label{eq:XPTHQFF1}
\end{equation}
Using the scaling in \cref{eq:scaleDC} one easily finds the scaling in \cref{eq:scaleFF}. The important point here is that the dominant physics of these form factors arises from the operator that involves $P_Q, P^{*}_Q$ and $\pi$. Since in our model, so far, we are missing $P^{*}_Q$, we aim to reproduce \cref{eq:XPTHQFF1} only in the limit $g_\pi \rightarrow 0$. It is, however, straight-forward to implement the heavy quark spin symmetry. We leave this for our future endeavors.    

\subsection{Leptonic and semileptonic decays of $B_c$ meson from heavy quark spin symmetry}
\label{subsec:hqet_heavyheavy}
The $B_c$ pseudoscalar meson, which decays only via the weak interaction (unlike other heavy quarkonium systems), uniquely probes both weak and strong interactions. For leptonic decays, all perturbative and nonperturbative QCD effects enter the decay rate through the decay constant $f_{B_c}$:  
\begin{equation}
    \label{eq:fBc}
    f_{_{\!B_c}} \ \equiv \ i \: \BState{0} \bar{b} \, \slashed{p}_{_{\!\!B_c}} \, 
                \gamma^5 c(0) \State{B_c\left(p_{_{\!B_c}} \right)} /m_{_{\!B_c}}^2 \; .
\end{equation} 
Non Relativistic Quantum Chromo Dynammics (NRQCD) \cite{Bodwin:1994jh,Lepage:1992tx} allows one to factorize the short- and long-distance effects involved in the calculation of the decay rate. One estimates short-distance factors using perturbative QCD and the long-distance matrix elements of NRQCD by using either the lattice or the wavefunction of the $B_c$ state \cite{Kiselev:1992tx,Kiselev:1992au,Kiselev:1995bv,Jenkins:1992nb}. Since we are interested in mostly scalings of the decay constant, we describe here the latter.   
\begin{equation}
    f_{_{\!B_c}} \ \sim \ \left( 1 + \Order{\frac{\alpha_s}{\pi}} \right) \frac{1}{\sqrt{m_{_{B_c}}}} \Psi\left(0\right) \ \sim \ \frac{1}{\sqrt{m_{B_c}}} \: a_0^{-3/2} + \ldots \; ,
    \label{eq:scaleDCBc}
\end{equation}
where $\Psi(x)$ and $a_0$ represent the wave function and the characteristic length scale of the $B_c$ state, respectively. In the limit $m_b \gg m_c \gg \Lambda$ one can use Coulombic wavefunctions for $B_c$, where $a_0$ is the Bohr radius, which scales with the inverse of the reduced mass (i.e., $a_0 \sim \left( m_c+m_b\right)/m_b m_c \sim 1/m_c$). 

Semileptonic decays, such as $B_c \rightarrow B_{d,s} \: \bar{\ell} \nu$, which go through $c \to \{ d,s\}$ transitions, are also important to us. Scalings of these rates with the meson masses and with the $a_0$ factor can be estimated following works by Jenkins et. al.~\cite{Jenkins:1992nb}. As we show shortly, these transitions become important, especially for disentangling scalings of a few of the Wilson coefficients that are under consideration here. The primary ansatz here is that the final state meson (call it $Y$) continues to be in the same velocity eigenstate as the initial state (call it $X$) after the decay $X \to Y \bar{\ell} \nu$ 
\begin{equation}
    p_{_{X}}^\mu \ \equiv \ m_{_{X}} v^\mu 
    \quad  \text{and} \quad \ 
    p_{_{Y}}^\mu \ \equiv \ m_{_{Y}} v^\mu + q^\mu \; ,  
\label{eq:vel_expn}
\end{equation}
$q^\mu$ being the small residual momentum of $Y$. The momentum imbalance (i.e., $p_{_{X}}^\mu - p_{_{Y}}^\mu$) is transferred to the lepton system. It is possible to calculate the form factors associated with these decays in the limit $B_c$ is considered point-like (i.e., $a_0 \ll 1/\Lambda$).  
In case of zero recoil, the applications of the heavy quark spin symmetries reveal that~\cite{Jenkins:1992nb}
\begin{equation}
    \BState{B_{a}, v, q} \bar{q_a} \gamma_\mu  c  \State{B_{c},v}  \approx  
       A_1 \sqrt{m_{_{\!B_c}} m_{_{\!B_a}} }  \left( f_{_{\!B_a}} \sqrt{m_{_{\!B_a} }} \right) \int \!\! d^3\!x \  e^{iq\cdot x} \Psi(x) \ \sim \   
       \ m_B \:  a_0^{3/2}  \; ,  
  \label{eq:Bc2BqFF}     
\end{equation} 
where $a$ subscript stands for light quarks $\{d,s\}$,  $A_1$ is a numerical constant, $a_0$ is the Bohr radius of the $B_c$ state (approximately scales with $ 1/m_c$). We have neglected terms of order $a_0^2 \bf{q}^2$, with $\bf{q}$ being the recoil 3-momenta\footnote{States where we explicitly put $v$, are normalized to the zero component of velocity, $v^0$ rather than to energy $E$ and hence carry a different dimension than their relativistically normalized counterparts.}.  In order to arrive in the final scaling in the equation above, we completely neglect the mass split between $B_c$ and $B_a$ and used the earlier result that $f_{B_a} \sqrt{m_{B_a}}$ does not scale (from \cref{eq:scaleDC}).  Note that the form factor also has a $a_0 q^\mu$ dependence in general, which does not contribute to the zero recoil. 

It is tempting to include a discussion of $B_c \rightarrow D^0 \: \bar{\ell} \nu$, (which proceeds via the $\bar{b}\to \bar{u}$ transition) or $B_c \rightarrow \eta_c \: \bar{\ell} \nu$, (via the $\bar{b}\to \bar{c}$ transition), based on the same logic as before. However, the applicability of the similar formalism is questionable, as the large mass-difference between $B_c$ and $D^0/\eta_c$  spoils the predictability of the niche velocity expansion. Note that the residual momentum $q^\mu$ becomes large (with respect to the scale $m_{c}$) in the limit $m_{_{B_c}} \gg m_{_{D^0}}$. Consequently, the amplitude receives large corrections from additional operators (of order $q/m_c$ ) signaling the breakdown of the expansion. However, in the opposing limit of $m_{_{B_c}} \simeq m_{_{D^0}}$, the formalism in \cref{eq:vel_expn} gives a well controlled expansion, and one can deduce a scaling identical to \cref{eq:Bc2BqFF}. Following the same recipe as before, one finds in the zero recoil 
\begin{equation}
    \BState{D^0, v, q} \bar{b} \gamma_\mu  u  \State{B_{c}, v}  \approx  
       A_2 \sqrt{m_{_{B_c}} m_{_{D^0}} }  \left( f_{_{D^0}} \sqrt{m_{_{D^0}}} \right)   \int \!\! d^3\!x \  e^{iq\cdot x} \Psi(x)  \sim    
        \sqrt{m_{{B}} m_{_{D}} }  \:  a_0^{3/2}  \; ,
  \label{eq:Bc2D0CFF}  
\end{equation}
where, $A_2$ is another numerical factor and we neglect terms of order $a_0^2 \bf{q}^2$ as before. The case for $B_c \rightarrow \eta_c \: \bar{\ell} \nu$ is slightly different, since there exists additional spin symmetry (because of $\bar{c}$ in the final state). One finds
\begin{equation}
    \BState{\eta_c, v, q} \bar{b} \gamma_\mu  c  \State{B_{c}, v}  \approx  
       A_3 \sqrt{m_{_{B_c}} m_{_{\eta_c}} }    \int \!\! d^3\!x \  e^{iq\cdot x} \Psi(x) \Psi^*_{\eta_c} \sim    
        \sqrt{m_{{B}} m_{\eta_c} }  \:  
            \frac{a_0^{3/2} a_{{\eta_c}}^{3/2}}
                {\left(a_0 + a_{\eta_c}\right)^3 } \; ,
  \label{eq:Bc2EtaCFF}  
\end{equation}
where $A_3$ is a numerical constant, $\Psi_{\eta_c}(x)$ and $a_{\eta_c}$ represent the wavefunction and the characteristic length scale of the state $\eta_c$, respectively. Note that the R.H.S. of \cref{eq:Bc2EtaCFF} can be simplified further by noting that in the limit $m_b \gg m_c$, both $a_0$ and $a_{\eta_c}$ scale identically with $m_c$ and hence the matrix element on the L.H.S. scales simply as   $\sqrt{m_{{B}} m_{\eta_c} }$.  

\subsection{Leptonic and Semileptonic decays of heavy mesons in our set-up}
\label{subsec:pheno_leptonic_model}

The purpose of this subsection is to identify operators that generate the leptonic and semileptonic transitions discussed in this section previously. Equating the amplitudes of these transitions in our set-up with those summarized before gives a nontrivial understanding of our Wilson Coefficients. 

We begin with the leptonic and semi-leptonic decays of the heavy-light mesons (i.e., $B^{+}$ and $D^{+}$). Of the operators listed in \cref{table:hadroniccurrents}, the following operators generate these decays at the leading order.  
\begin{equation}
\begin{split}
     \left(  \mathcal{K}_5 \Oph_5 +   \mathcal{K}_8 \Oph_8  \right) \Opl \ \supset & \
        V_{\bar{d}c} \Bigg[  \left( -\mathcal{K}_8 \Lambda_c 
                + \frac{1}{2} \mathcal{K}_5 f_\pi \right) \partial_\mu D^{+} 
                        + \frac{i\mathcal{K}_5 }{\sqrt{2}}\ \pi^0  \darrowD D^{+} 
                    \Bigg] \Opl\;,  \\
    \left(  \mathcal{K}_6 \Oph_6 +   \mathcal{K}_7 \Oph_7  \right) \Opl  \ \supset & \
        V_{\bar{b}u} \Bigg[  \left(  \mathcal{K}_6 \Lambda_b 
            - \frac{1}{2} \mathcal{K}_7 f_\pi \right) \partial_\mu B^{+} 
                - \frac{i\mathcal{K}_7}{\sqrt{2}}\  \pi^0  \darrowD B^{+} 
                    \Bigg] \Opl \; .  
\end{split}
\label{eq:lepdecayops}
\end{equation}
It is straightforward to read off the decay constants and form factors from \cref{eq:lepdecayops}.  
For example, one finds the amplitude for leptonic decays to be:
\begin{equation}
\begin{split}   
        \mathcal{M}\left(D^+ \to \bar{\ell}\nu\right) \ & = \ \frac{4G_F}{\sqrt{2}}\; V_{\bar{d}c}\; i\left(-\frac{1}{2} \mathcal{K}_{5} f_\pi+\mathcal{K}_{8}\Lambda_c\right) p_{D^+}^\mu\; \mathcal{A}_\mu^L  \; , \\
    \mathcal{M}\left(B^+ \to \bar{\ell}\nu\right) \ & = \ \frac{4G_F}{\sqrt{2}}\; V_{\bar{b}u}\; i\left(\frac{1}{2} \mathcal{K}_{7} f_\pi-\mathcal{K}_6 \Lambda_b\right) p_{B^+}^\mu\; 
        \mathcal{A}_\mu^L\;  ,
\end{split}    
\end{equation}
where $\mathcal{A}_\mu^L$ represents the part of the amplitude that involves leptons only. This gives us expressions for decay constants in our model at leading order:    
\begin{equation}
    f_D \ = \ \mathcal{K}_5 f_\pi-2\mathcal{K}_{8}\Lambda_c \qquad  \text{and} \qquad
    f_B \ = \ -\mathcal{K}_{7} f_\pi+2\mathcal{K}_6 \Lambda_b \; . 
\label{eq:derDC}
\end{equation}
For our model to faithfully reproduce the phenomenology of SM hadrons, it must have the right scalings, which we mentioned at the beginning of this section. This suggests that if we utilize \cref{eq:scaleDC} as input, we can estimate the scalings of the Wilson coefficients
\begin{equation}
    \mathcal{K}_5 f_\pi - 2 \mathcal{K}_{8}\Lambda_c \sim \mathcal{O}\left( \frac{1}{\sqrt{m_D}}\right) 
        \quad \text{and} \quad
    \mathcal{K}_7 f_\pi - 2  \mathcal{K}_{6}\Lambda_b \sim \mathcal{O}\left( \frac{1}{\sqrt{m_B}}\right) \; .
\label{eq:derScalingsDC}    
\end{equation}
In case the terms in \cref{eq:derDC} are saturated by $\mathcal{K}_5$ and $\mathcal{K}_7 $, respectively, one gets a neat scaling of $\mathcal{K}_5/\mathcal{K}_7$. 

It is a little more work to figure out the prediction of form factors in our scheme since we should be able to faithfully reproduce SM leading order results only in the NR limit. The first job is to go to the NR limit of the operators listed in \cref{eq:lepdecayops} and consider the terms leading in $m_P$ to write down the amplitude.    
\begin{equation}
    \begin{split}
           \BState{\pi^0} \mathcal{K}_5\mathcal{O}_5^h + \mathcal{K}_8\mathcal{O}_8^h \State{D^+} \ & = \ 
                i  V_{\bar{d}c}\left(\frac{\mathcal{K}_5}{\sqrt{2}}\right)
                   m_D v^\mu \; , \\
           \BState{\pi^0} \mathcal{K}_6\mathcal{O}_6^h + \mathcal{K}_7\mathcal{O}_7^h \State{B^+} \ & = \ 
                i V_{\bar{b}u}\left(\frac{-\mathcal{K}_7}{\sqrt{2}}\right)
                    m_B v^\mu \; , \\
            \implies \qquad  
        f_{+}^{D \rightarrow \pi} + f_{-}^{D \rightarrow \pi}  = \frac{1}{\sqrt{2}} \mathcal{K}_5 \; ,   \qquad & \qquad  
        f_{+}^{B \rightarrow \pi} + f_{-    }^{B \rightarrow \pi} = - \frac{1}{\sqrt{2}} \mathcal{K}_7  \;, \\  
         \text{and} \quad 
        f_{+}^{B/D \rightarrow \pi} & - f_{-}^{B/D \rightarrow \pi}  = 0 \; . 
    \end{split}
\end{equation}
This reproduces \cref{eq:XPTHQFF1} in the $g_\pi \rightarrow 0$ limit as claimed earlier. Furthermore, if one takes into consideration the size of $\mathcal{K}_5$ and $\mathcal{K}_7$ from \cref{eq:derScalingsDC}, we get the desired scaling of the form factors for semi-leptonic decays, suggesting that both the terms in \cref{eq:derDC} are indeed saturated by $\mathcal{K}_5$ and $\mathcal{K}_7 $, respectively.  

We next discuss various (semi)leptonic decays of $B_c$. As before, we essentially identify operators that cause these transitions and then utilize \cref{eq:Bc2BqFF} to learn about the corresponding Wilson coefficients and the phenomenological parameters $\Lambda_{b,c}$. Note that the operators in question must involve at least one factor of $\Sigma_H$ (that is, $\Oph_2, \Oph_6$ and $\Oph_8$), which contains the field $B_c$.    
\begin{equation} 
\begin{split}
    \mathcal{K}_2 \Oph_2 \   \supset  \  
                V_{\bar{b}c} \mathcal{K}_2 \Bigg[  i\left(  \Lambda_c + \Lambda_b  \right) \partial_\mu B_c^{+} \ - \  
                         \eta_c  \darrowD B_c^{+} 
                    \Bigg]  \; , \\
    \mathcal{K}_6 \Oph_6 \   \supset  \  
               i V_{\bar{b}u} \mathcal{K}_6  \
                         \bar{D}  \darrowD B_c^{+} 
                     \; , 
                    \quad \text{and} \quad             
        \mathcal{K}_8 \Oph_8 \  \supset  \  
               i V_{\bar{s}c}  \mathcal{K}_8 \
                          \bar{B_s}  \darrowD B_c^{+} 
                     \; . 
\end{split}
\end{equation}
As usual, we can read off the amplitudes 
\begin{align}
     \mathcal{M}\left( B_c^+ \to \bar{\ell}\nu \right) \ &= \ \frac{4G_F}{\sqrt{2}}V_{\bar{b}c}\;\mathcal{K}_2 
        \left(\Lambda_b \ + \ \Lambda_c\right) p^\mu_{B_C} \mathcal{A}_\mu^L\;,
        \label{eq:ModelDCBc} \\
     \mathcal{M}\left( B_c^+ \to B_s \, \bar{\ell} \nu \right) \ &= \ \frac{4G_F}{\sqrt{2}} V_{\bar{s}c}\; \mathcal{K}_8 \left(p^\mu_{B_c}+p^\mu_{B_s}\right)\mathcal{A}_\mu^L\; ,
     \label{eq:Bc2BsFFModel}\\
          \mathcal{M}\left( B_c^+ \to D \, \bar{\ell} \nu \right) \ &= \ \frac{4G_F}{\sqrt{2}} V_{\bar{b}u}\; \mathcal{K}_6 \left(p^\mu_{B_c}+p^\mu_{D}\right)\mathcal{A}_\mu^L\;,
     \label{eq:Bc2DFFModel}\\
    \mathcal{M}\left( B_c^+ \to \eta_c \, \bar{\ell} \nu \right) \ &= \ \frac{4G_F}{\sqrt{2}} V_{\bar{b}c}\; i\mathcal{K}_2 \left(p^\mu_{B_c}+p^\mu_{\eta_c}\right)\mathcal{A}_\mu^L\;   .  
     \label{eq:Bc2EtaFFModel}
\end{align}

We can now systematically compare the various terms in \cref{eq:Bc2BsFFModel} with those stated in the last subsection. Let us begin with the fully leptonic channel for the decays of $B_c$. It is clear from \cref{eq:scaleDCBc}, that for the amplitude in \cref{eq:ModelDCBc} to have the same scaling, the combination $\Lambda_b + \Lambda_c$ must depend only on $a_0$. A satisfying ansatz is 
\begin{equation}
        \Lambda_{c} \ = \ \frac{\lambda}{m_c}  
            \quad \text{and} \quad 
    \Lambda_{b} \ = \ \frac{\lambda}{m_b} \; ,
    \qquad 
    \implies  \Lambda_{c} + \Lambda_{b} \ = \   \lambda 
        \left( \frac{1}{m_c} + \frac{1}{m_b} \right) 
        \ \sim \ \lambda a_0  \; , 
\label{eq:Ansatz1Lambda}
\end{equation}
where $\lambda$ is a common factor, which can in general be functions of heavy quark/meson masses (or may even include factors of $a_0$ and $a_{\eta_c}$). This ansatz also determines the scaling of $\mathcal{K}_2$ itself with the ambiguity of the factor $\lambda$. 
\begin{equation}
     \mathcal{K}_2 \: \lambda \ \sim \ \frac{1}{\sqrt{m_B^{\phantom{1}} }} \:  a_0^{-5/2}   \; . 
\label{eq:Ansatz2Lambda}    
\end{equation}
We can now use the final piece, namely the scaling of the matrix element in decay $B_c \to \eta_c$ in \cref{eq:Bc2EtaCFF} and compare the amplitude in \cref{eq:Bc2EtaFFModel} to arrive at the following. 
\begin{equation}
    \mathcal{K}_2  \ \sim \ \frac{\sqrt{m_B^{\phantom{1} } \: m_{\eta_c}}}{m_B + m_{\eta_c}} \ \sim \ \sqrt{ \frac{m_{\eta_c}^{\phantom{1}}}{m_B^{\phantom{1} } } }  \; , 
    \quad \implies \quad 
    \lambda \ \sim \ \frac{1}{\sqrt{m_{\eta_c}^{\phantom{1}} }} \:  a_0^{-5/2}  
    \ \sim \ a_0^{-2} \; ,
\end{equation}
where we have used the fact that both $1/a_0$ and $m_{\eta_c}$ have the same scaling w.r.t. the charm mass parameter $m_c$ in the limit $m_b \gg m_c$. 

It is straightforward to see that a similar comparison of the matrix element in \cref{eq:Bc2BqFF} with the amplitude in  \cref{eq:Bc2BsFFModel} yields non-trivial results as well. Neglecting the mass splitting between $m_{B_c}$ and $m_{B_q}$  in the form factor, we realize that $\mathcal{K}_8$ does not scale with $m_B$ but scales with $a_0$.
\begin{equation}
    \mathcal{K}_8 \ \sim \ a_0^{3/2} \qquad \implies \qquad  
    \mathcal{K}_8 \Lambda_c \ \sim \ a_0^{3/2} \frac{1}{m_c} a_0^{-2} 
        \ \sim \ \frac{1}{\sqrt{m_c}} \ \sim \ \frac{1}{\sqrt{m_D}} \; ,
    \label{eq:scalingK8Lc}
\end{equation}      
where we use the fact that $m_D$ scales with $m_c$. Therefore, we find that both $\mathcal{K}_5$ and $\mathcal{K}_8 \Lambda_c $ scale in the same way as dictated by \cref{eq:derDC}.  

It is tempting to compare the amplitudes for $B_c \to D$ in \cref{eq:Bc2DFFModel} with the matrix element in \cref{eq:Bc2D0CFF} as well, which, as mentioned before, are more appropriate if one neglects the mass difference $m_{B_c} - m_{D}$ and takes the zero recoil limit. In any case, just as with the $B_c \to \eta_c$ transitions, we proceed with the assumptions that the scalings shown in \cref{eq:Bc2D0CFF} still hold.  
\begin{equation}
\begin{split}
    \mathcal{K}_6  \ \sim \ \frac{\sqrt{m_{B_c}^{\phantom{1}} m_{D}^{\phantom{1}}} }{ m_{B_c}^{\phantom{1}} + m_{D}^{\phantom{1}} } \: a_0^{3/2} 
    \ \sim \ \sqrt{ \frac{m_{D}^{\phantom{1}}} { m_{B_c}^{\phantom{1}} } } \: a_0^{3/2} 
     \qquad  \implies \qquad  
    \mathcal{K}_6 \Lambda_b  \ \sim \ \left( \frac{m_D}{m_B} \right) \frac{1}{\sqrt{m_B}} \; .
\end{split}        
    \label{eq:scalingK6}
\end{equation}  
We again use a similar argument to that before to arrive at \cref{eq:scalingK6}. Comparing it with $f_B$, the $B$ meson decay constant as given in \cref{eq:derDC}, we conclude that if the scalings of \cref{eq:Bc2D0CFF}, as derived from heavy quark spin symmetry, have to hold, $f_B$ must be saturated by $\mathcal{K}_7$, since $\mathcal{K}_6 \Lambda_b$ is suppressed by an extra factor of $m_D/m_B$.

\section{Consistency checks in hadronic decays via isospin sum rules}
\label{sec:consistency_hadronic}

The purpose of this section is to demonstrate that the well-established amplitude relations among hadronic decays $B\to \pi_i \pi_j$, where $\pi_{i,j}$ being the pNGBs, are automatic in our phenomenological constructions. Given that these relations are proposed using arguments based on considerations from vectorial subgroup $SU(3)_{V}$ of the chiral symmetry $SU(3)_L\times SU(3)_R$, which our set-up employ maximally this should not come at a surprise. Yet, these provide additional consistency checks for the full model which break the symmetry pattern using the electroweak charge matrices as spurions. 

We begin this section with the conventional understanding of these relations. To begin, consider all the pnGBs except for $\eta_0$ in \cref{eq:defCompFieldsSigma}, which transforms as the adjoint of $SU(3)_{V}$ -- this is the representation often referred to as the pion octet. The final state consisting of two octets can be decomposed into irreducible representations of $SU(3)_{V}$ itself.

In practice, these reduced matrix elements are often re-expressed in terms of topological amplitudes such as tree ($T,T^\prime$), color-suppressed ($C,C^\prime$), and penguin ($P,P^\prime$) etc., where the prime indicates strangeness changing processes ($q\in s$). This admits a more direct interpretation in terms of quark-level processes. The tree-level operators $\bar b \to \bar q\, u \bar u$ transform as $\bar{3}\otimes 3\otimes\bar{3}=\bar{3}\oplus\bar{3} \oplus 6 \oplus \overline{15}$, under $SU(3)_F$, where $b$ is a spectator singlet. The $\overline{15}$ component being symmetric in flavor indices, provide the dominant contribution to the 27-plet in the final state since $\langle (PP)_{1,8,27} \,|\, H_{\text{eff}} \,|\, B(\bar 3)\rangle$~\cite{Zeppenfeld:1980ex,Savage:1989ub,Chau:1990ay,Grinstein:1996us}  . Consequently, the tree amplitudes $T,T^\prime$ and $C,C^\prime$ are closely related to this symmetric component. On the other hand, penguin operators, which predominantly transform as $\bar{3}$, naturally project onto singlet and octet structures. Thus, the topological amplitudes can be understood as specific linear combinations of the reduced matrix elements associated with the 1, 8 and 27 representations. We now discuss these topological amplitudes in some detail~\cite{Gronau:1994rj,Gronau:1995hm,Gronau:1995hn}.

\begin{itemize}
\item \textbf{Tree} ($T, T^\prime$) or color-favored: At the quark level, this corresponds to the transition $\bar{b} \to \bar{q} u \bar{u}$. In this case, the $\bar{q} u$ system forms a color-singlet pseudoscalar meson, while the $\bar{u}$ combines with the spectator quark to form the second meson.
\item \textbf{Color-suppressed} ($C, C^\prime$): Here, the $\bar{u}u$ system hadronizes into a pseudoscalar meson, whereas the $\bar{q}$ combines with the spectator quark to form the other meson.
\item \textbf{Penguin} ($P, P^\prime$): These amplitudes arise from loop-induced transitions $\bar{b} \to \bar{q}$, mediated by gluon, electroweak, or other heavy particle exchanges.
\item \textbf{Exchange} ($E, E^\prime$): In this topology, the $\bar{b}$ quark and the $q$ quark exchange a $W$ boson in the $t$ channel, producing a $u \bar{u}$ final state. The spectator quark subsequently participates in the formation of the two pseudoscalar mesons in the final state.
\item \textbf{Annihilation} ($A, A^\prime$): This is a direct annihilation process $\bar{b} u \to \bar{q} u$ mediated by the exchange $W$, with both initial-state quarks participating in the weak interaction.
\item \textbf{Penguin annihilation} ($PA, PA^\prime$): In this case, the initial $\bar{b} q$ state annihilates to vacuum quantum numbers via a penguin loop, subsequently hadronizes into two pseudoscalar mesons.
\end{itemize}
As noted in~\cite{Gronau:1994rj}, the amplitude structure for the processes $B \to K\pi$ is given in Table~\ref{tab:example}:
\begin{table}[h!]
\centering
\begin{tabular}{|c|c|c|c|c|c|c|}
\hline
Processes & $T^\prime$ & $C^\prime$  & $P^\prime$ & $E^\prime$ & $A^\prime$ & $PA^\prime$ \\
\hline
$B^+\to K^+\pi^0$ & $-\frac{1}{\sqrt{2}}$ & $-\frac{1}{\sqrt{2}}$ & -$\frac{1}{\sqrt{2}}$ & 0 & -$\frac{1}{\sqrt{2}}$ & 0 \\
\hline
$B^0\to K^0\pi^0$ & 0 & -$\frac{1}{\sqrt{2}}$ & $\frac{1}{\sqrt{2}}$ & 0 & 0 & 0 \\
\hline
$B^+\to K^0\pi^+$ & 0 & 0 & 1 & 0 & 1 & 0 \\
\hline
$B^0\to K^+\pi^-$ & -1 & 0 & -1 &0 & 0 & 0 \\
\hline
\end{tabular}
\caption{Coefficients for $B\to K\pi$ process.}
\label{tab:example}
\end{table}

Notice that there is a freedom to choose the overall sign of different topological processes. The particular choice as shown in Table~\ref{tab:example} leads to
\begin{equation}
\begin{split}
\sqrt{2}\: \mathcal{M}_{B^+\to K^+\pi^0}
\ + \ \mathcal{M}_{B^+\to K^0\pi^+}
\ - \  \sqrt{2} \: \mathcal{M}_{B^0\to K^0\pi^0} 
\ - \  \mathcal{M}_{B^0\to  K^+\pi^-} \ = \ 0 \;.
\label{eq:sum}
\end{split}
\end{equation}
Furthermore, each sum in Eq.~\eqref{eq:sum} corresponds to the combination $T^\prime + C^\prime$, that is, the tree and color suppressed contributions. 

In our framework, the operators resulting in the process $B \to K\pi$ should exactly contain a single factor of the field $H$ and at least one factor of the field $\Sigma$ for the pNGBs. Given that $\Sigma_H$ contains numbers (that is, $\Lambda_{b,c}$) in its diagonal positions, the operator can also contain one or more of $\Sigma_H$. These restrictions significantly narrow the set of operators relevant to this analysis. In particular, we find four operators (two double trace operators and two single trace operators) that contribute to $B \to K\pi$. 
\begin{equation}
    \mathcal{C}_1^6 \: {\Oph_1}^{\dagger} \Oph_6 \ + \ 
        \mathcal{C}_1^7 \: {\Oph_1}^{\dagger} \Oph_7 \ + \ 
        \mathcal{E}_{19} \: \Ophh_{19} \ + \ \mathcal{E}_{38} \: \Ophh_{38}\;.
\label{eq:ops4B2PP}
\end{equation}

We summarize the amplitudes for each process in Table~\ref{table:BKpiamplitudes}. In particular, look at $\State{B^+} \to \State{K^+ \pi^0}$ to be specific. It receives contributions from all four operators in \cref{eq:ops4B2PP}. The piece proportional to $\mathcal{C}^{7}_1$ is the most straightforward to visualize with perturbative Feynman diagrams characterized by the "Tree" topology. Other contributions are expected to be sub-leading and require gluon exchanges and/or loop(s) in the quark picture. For example, the term proportional to $\mathcal{E}_{38}$ arises from diagrams with both valence quarks in $\State{B^+}$ participating, need an extra gluon exchange, and is color suppressed. Pieces involving $\mathcal{C}^{6}_1$ and $\mathcal{E}_{19}$ require insertions of $\langle \bar{b}b \rangle$ and $\langle \bar{c}c \rangle$ condensates respectively. An example diagram that may yield an amplitude similar to the piece proportional to $\mathcal{E}_{19}$ involves a penguin with the charm quark in the loop and additional gluon exchanges. Not surprisingly, this contribution is proportional to $V_{\bar{b}c}V_{\bar{c}s}$.         

\begin{table}[ht!]
\renewcommand{\arraystretch}{1.2} % slightly more vertical space
\centering
\begin{tabularx}{\linewidth}{|p{3cm}|X|}
\hline
\textbf{Transitions} & \textbf{Amplitudes} \\
\hline
\[\State{B^0} \to \State{K^0 \pi^0}\] &
\[
\begin{aligned}
\frac{4G_F}{\sqrt{2}}\, \Bigg[&V_{\bar{b}u} V_{\bar{u}s}\,f_\pi\Bigg\{\mathcal{E}_{38}\frac{\left(3m_B^2-2m_K^2\right)}{4\sqrt{2}}-\mathcal{C}^{7}_1\frac{\left(2m_B^2-m_K^2-m_\pi^2\right)}{2\sqrt{2}}\\
&+\mathcal{C}^{6}_1\frac{\Lambda_b}{f_\pi}\frac{\left(m_\pi^2-m_K^2\right)}{\sqrt{2}}\Bigg\}+V_{\bar{b}c}V_{\bar{c}s}\;\mathcal{E}_{19}\frac{\Lambda_c^2}{\sqrt{2}f_\pi} m_B^2\Bigg] 
\end{aligned}
\]
\normalsize
\\
\hline
  \[ \State{B^+} \to \State{\pi^+ K^0} \] &
%\footnotesize
\[
\begin{aligned}
\frac{4G_F}{\sqrt{2}}\, \Bigg[&V_{\bar{b}u} V_{\bar{u}s}\,f_\pi\Bigg\{\mathcal{E}_{38}\frac{m_B^2}{4}+\mathcal{C}^{7}_1\frac{\left(m_K^2-m_\pi^2\right)}{2}
\\
&-\mathcal{C}^{6}_1\frac{\Lambda_b}{f_\pi}\left(m_K^2-m_\pi^2\right)\Bigg\}
+V_{\bar{b}c}V_{\bar{c}s}\;\mathcal{E}_{19}\frac{\Lambda_c^2}{f_\pi} m_B^2\Bigg] 
\end{aligned}
\]
\normalsize
\\
\hline
\[ \State{B^0} \to \State{K^+ \pi^-} \] &
%\footnotesize
\[
\begin{aligned}
\frac{4G_F}{\sqrt{2}}\, \Bigg[&V_{\bar{b}u} V_{\bar{u}s}\,f_\pi\Bigg\{\mathcal{E}_{38}\frac{m_B^2}{4}-\mathcal{C}^{7}_1\left(m_B^2-m_\pi^2\right)\Bigg\}\\
&+V_{\bar{b}c}V_{\bar{c}s}\;\mathcal{E}_{19}\frac{\Lambda_c^2}{f_\pi} m_B^2\Bigg] 
\end{aligned}
\] 
\normalsize
\\
\hline
\[ \State{B^0} \to \State{K^0 \pi^0} \] &
%\footnotesize
\[
\begin{aligned}
\frac{4G_F}{\sqrt{2}}\, \Bigg[&V_{\bar{b}u} V_{\bar{u}s}\,f_\pi\mathcal{E}_{38}\frac{\left(m_B^2-2m_K^2\right)}{4\sqrt{2}}-V_{\bar{b}c}V_{\bar{c}s}\;\mathcal{E}_{19}\frac{\Lambda_c^2}{f_\pi} \frac{m_B^2}{\sqrt{2}}\Bigg] 
\end{aligned}
\] 
\normalsize
\\
\hline
\end{tabularx}
\caption{Amplitudes for $\State{B}\to \State{K\pi}$ transitions.}
\label{table:BKpiamplitudes}
\end{table}

It is straightforward to verify that the amplitudes presented in Table~\ref{table:BKpiamplitudes} satisfy a sum rule analogous to that given in Eq.~\eqref{eq:sum}, with the only difference being an overall sign in the term $\mathcal{M}_{B^+\to K^0\pi^+}$. Importantly, our results are consistent with \cite{Wang:2024tnx}, but differ by a sign from those reported in \cite{Gronau:1990ka,Gronau:1994rj,Nir:1991cu,Lipkin:1991st}. This discrepancy can be traced to the convention adopted for the pion field, specifically for the identification of $\pi^- $. Using a field redefinition $\pi^- \to -\pi^-$, the corresponding term in the Lagrangian changes sign, which in turn flips the sign of the amplitude for $B^+ \to K^0\pi^+$. With this adjustment, our results come into agreement with the references aforementioned.

We take a similar exercise for the $|B\rangle \to |D\pi\rangle$ decays. We the operators that play an important role in these transitions are
\begin{equation}
\mathcal{C}_1^3\;\mathcal{O}_1^{h\dagger}\mathcal{O}_3+\mathcal{E}_{15}\;\mathcal{O}_{15}^{hh}+\mathcal{E}_{22}\;\mathcal{O}_{22}^{hh}+\mathcal{E}_{23}\; \mathcal{O}^{hh}_{23}\;.
\end{equation}
Take, for example, the process $|B^0\rangle\to|D^{-}\pi^+\rangle$. The double-trace operator can be easily understood from the fact that both the $B$ and $D$ mesons are embedded in the field $H$, and therefore the operator $\mathcal{O}_3$ contributes. As in the previous example, this contribution can be identified with the tree-level topology. The same quark-level process can also proceed via an additional gluon exchange, leading to contractions of color indices between different quark lines. This gives rise to a color-suppressed contribution, which is associated with the operator $\mathcal{O}_{23}$. The contribution of the operator $\mathcal{O}_{22}$ requires an insertion of $\langle b\bar{b}\rangle$ and is therefore suppressed. We present the corresponding amplitudes in Table~\ref{table:BDpiamplitudes}.
\label{appendix:1}
\begin{table}[ht!]
\renewcommand{\arraystretch}{1.2} % slightly more vertical space
\centering
\begin{tabularx}{\linewidth}{|p{3cm}|X|}
\hline
\textbf{Transitions} & \textbf{Amplitudes} \\
\hline
%\centering 
\[\State{B^0} \to \State{D^- \pi^+}\] &
%\footnotesize
\[
\begin{aligned}
i\frac{4G_F}{\sqrt{2}}\, &V_{\bar{b}c} V_{\bar{u}d}\,\Bigg[f_\pi\left(\mathcal{C}_1^{3}-\frac{\mathcal{E}_{23}}{2}\right)\left(m_B^2-m_D^2\right)-\Lambda_b \mathcal{E}_{22}\left(m_B^2-m_\pi^2\right)\Bigg]
\end{aligned}
\]
\normalsize
\\
\hline
\[\State{B^0} \to \State{\bar{D} \pi^0}\] &
%\footnotesize
\[
\begin{aligned}
i\frac{4G_F}{\sqrt{2}}\, &V_{\bar{b}c} V_{\bar{u}d}\,\Bigg[\Lambda_c\mathcal{E}_{15}\frac{\left(m_\pi^2-m_D^2\right)}{\sqrt{2}}-\Lambda_b\mathcal{E}_{22}\frac{\left(m_B^2-m_\pi^2\right)}{\sqrt{2}}\Bigg]
\end{aligned}
\]
\normalsize
\\
\hline
\[\State{B^+} \to \State{\bar{D} \pi^+}\] &
%\footnotesize
\[
\begin{aligned}
i\frac{4G_F}{\sqrt{2}}\,V_{\bar{b}c} V_{\bar{u}d}\, \Bigg[f_\pi\left(\mathcal{C}_1^3-\frac{\mathcal{E}_{23}}{2}\right)\left(m_B^2-m_D^2\right)+\Lambda_c \mathcal{E}_{15}\left(m_D^2-m_\pi^2\right)\Bigg] 
\end{aligned}
\] 
\normalsize
\\
\hline
\end{tabularx}
\caption{Amplitudes for $|B\rangle \to |D\pi\rangle$ transitions.}
\label{table:BDpiamplitudes}
\end{table}
This leads to the following sum rule.
\begin{equation}
    \mathcal{M}\left(B^0\to D^-\pi^+\right) - \sqrt{2} \;\mathcal{M}\left(B^0\to \bar{D}\pi^0\right) = \mathcal{M}\left(B^+\to D\pi^+\right)\;.
\end{equation}
Similarly to the previous case, notice again the difference in sign with~\cite{Chiang:2002tv,Fayyazuddin:2004ac,Pirjol:2004yf}.
%----------------------------
\section{Phenomenology of \texorpdfstring{$B\to K + \eta_c/\eta'/\eta$}{B2KY} }
\label{sec:B2KY}
%----------------------------
The purpose of this section is to discuss the process of $B\rightarrow K+Y$ where $Y$ stands for any of the neutral mesons, $\eta, \eta'$ and $\eta_c$. As mentioned in the Introduction, the observed hierarchy in these branching ratios is difficult to explain using quark level diagrams, and it has received a lot of attention over the years. To be specific, PDG~\cite{ParticleDataGroup:2024cfk} gives the branching fractions to be   
\begin{equation}
\frac{\BR(B^+\to K^+\eta_c)}{\BR(B^+\to K^+\eta^\prime)} \sim 15  \qquad  \text{and} \qquad \frac{\BR(B^+\to K^+\eta^\prime)}{\BR(B^+\to K^+\eta)} \sim 29\;.  
\label{eq:B2KYratios}
\end{equation}
These observations in \cref{eq:B2KYratios} necessitate understanding beyond the naive quark model pictures. Although the decay $B^+\to K^+\eta_c$ is CKM enhanced with respect to $B^+\to K^+\eta/\eta'$, it is, nevertheless, color suppressed. In addition, a rather large rate of $B^+\to K^+\eta'$ when compared to $B^+\to K^+\eta$ is not expected. It is straightforward to conclude that the usual description of these modes in terms of the $b$-quark decays to light quarks is not sufficient. 

The same conclusion can be reached via a more systematic reasoning. Consider, for example, the $b\to s$ quark-level effective Hamiltonian responsible for the $B^+\to K^+\eta^{(\prime)}$ decays 
\begin{equation}
    \mathcal{H}_{\text{eff}} = \frac{G_F}{\sqrt{2}}\: 
    \bigg[ \sum_{q = u,c} V_{\bar{b}q} V_{\bar{q}s} 
        \left( C_1^q O_1^q + C_2^q O_2^q \right)  - V_{\bar{b}t}V_{\bar{t}s} \sum_{i=3}^{12} C_i O_i \bigg]\;,
\end{equation}   
where $O_1^q = \left(\bar{q}_{\alpha} b_{\alpha} \right)_{_{V-A}} (\bar{s}_{\beta}q_{\beta})_{_{V-A}}$ and $O_2^q = (\bar{q}_{\beta}b_{\alpha})_{_{V-A}} (\bar{s}_{\alpha}q_{\beta})_{_{V-A}}$, with $q=u,\,c$ and $\alpha, \beta$ represent spinor indices. The operators $O_i$ include the QCD penguin operators (for $i=3$ to $i=6$), the electroweak penguin operators (for $i=7$ to $i=10$) and the electromagnetic and QCD dipole operators ($O_{11}$ and $O_{12}$ respectively). Of these, the important Wilson coefficients at the scale $\mu \sim m_b$ have been found to be~\cite{Buchalla:1995vs}
\begin{equation}
   C_1 \sim 1.1 \; , \quad  C_2 \sim -0.17 \; , \quad  
   C_{11}^{\text{eff}} \sim -0.3 \; , \quad\text{and} \quad C_{12}^{\text{eff}} \sim -0.14 \; ,
\end{equation}
where we have taken $C_{1,2} = C_{1,2}^u = C_{1,2}^c$. The other Wilson coefficients are one to two orders smaller and will be ignored. In fact, it suffices to focus on $O_{1,2}^q$ for the present discussion. The chromo magnetic dipole operator may be enhanced in some new physics models and can contribute to $B^+\to K^+\eta^\prime$ via the coupling of the gluon with $\eta^\prime$. However, in the SM its contribution is subdominant, and we do not discuss this possibility here. Note also that it is straightforward to include these neglected operators.

Working in the factorization approach, the leading part of the amplitude (matrix element) for the mode $B^+\to K^+\eta^\prime$, therefore, reads
\begin{equation}
\begin{split}
        \mathcal{M}(B^+\to K^+\eta^\prime) \ = \  \frac{G_F} {\sqrt{2}} \: 
     V_{\bar{b}u} V_{\bar{u}s} \bigg[ 
        a_1 \underbrace{\BState{ \eta'} \left( \bar{u}b \right)_{_{V-A}} 
            \State{B^+}}_{F_0^{B\to \eta^\prime}(m_{K}^2)} \: 
            \underbrace{\BState{ K^+}\left(\bar{s}u \right)_{_{V-A}} \State{0} }_{f_K}  \\
         \ + \ a_2  \underbrace{\BState{ \eta'} \left( \bar{u}u \right)_{_{V-A}} 
            \State{0}}_{f_{\eta'}^u} \: 
            \underbrace{\BState{ K^+}\left(\bar{s}b \right)_{_{V-A}} \State{B^+} }_{F_0^{B\to K}(m_{\eta'}^2)}  
     \bigg]    \ + \ u \rightarrow c \; .  
\end{split}    
\label{eq:MEB2KY}
\end{equation}
In this equation, $F_0$ is the $B\to P$ form factor and $f_K$ denotes the decay constant for $K$. The quantity $f_{\eta^\prime}^u$ denotes the up-quark contribution to the $\eta^\prime$ decay constant (defined via $\BState{\eta'} (\bar{q}q)_{_{V-A}}\State{0} \propto {f}^q_{\eta'}$). 

At this point, the naive expectation suggests that one can neglect the matrix elements of operators with charm quarks, since they must be suppressed in the constituent-quark model. However, it is straightforward to check that if one neglects the charm contribution in \cref{eq:MEB2KY}, the branching fraction computed turns out to be almost two orders of magnitude smaller than the experimentally observed one. 

One often cites the Okubo-Zweig-Iizuka (OZI) rule (also routinely simply referred to as the Zweig rule) ~\cite{Okubo:1963fa,Zweig:1964jf,Iizuka:1966fk} violating concept of ``intrinsic charm content of $\eta'$" to explain this large rate\footnote{This concept relies on the prominent role of gluons $\rightarrow \bar{c}c$ transition and, therefore, end up violating the Zweig rule.}. This contribution can be thought of as arising from the anomaly or through mixing $\eta_c$ - $\eta^\prime$. In either way, this suggests that the charm contribution in \cref{eq:MEB2KY} can not be neglected -- in fact, they might even dominate. 
In the factorization picture (\cref{eq:MEB2KY}), the effect of the charm contribution to the desired matrix element is proportional to the charm contribution to the $\eta'$ decay constant:  
\begin{equation}
    \mathcal{M}(B^+\to K^+\eta^\prime) \ \supset \   V_{\bar{b}u} V_{\bar{u}s}  [..]\: f_{\eta^\prime}^u \ + \  V_{\bar{b}c} V_{\bar{c}s}  (..) \: f_{\eta'}^c \ + \ \ldots \;,
\label{eq:eq:MEB2KYwC}
\end{equation}
where $\ldots$ represent various numerical factors as well as other terms. 

%----------------------------

\subsection{\texorpdfstring{$B\to K + \eta_c/\eta'/\eta$ }{} in our set-up}
\label{subsec:B2KY_model}

Since the fundamental degrees of freedom (i.e. the fields) in our set-up contain the creation and annihilation operators of asymptotic meson states, including effects such as mixing of these states becomes rather straightforward and completely tractable, as we demonstrate in this section. 

In our phenomenological approach, we regard the states in $Y$ as flavor eigenstates, but not mass eigenstates. While mass-mixing between $\eta$ and $\eta'$ is fairly standard and part of course book material, the mixing of $\eta$ and $\eta'$ with $\eta_c$ is not that trivial. One can argue phenomenologically that the $B\rightarrow K+Y$ data itself point towards such mixings. However, as mentioned in the Introduction, our starting point is somewhat different. Motivated by the AMSB+SQCD limit of QCD Ref.~\cite{Csaki:future}, we expect, for example, the mixing angle of $\eta'$ with $\eta_c$ to be nonzero, vanishing only in the infinite heavy-quark mass limit. 

To be specific, we begin with the mass terms in the flavor basis where we explicitly write down off-diagonal components of the squared-mass matrix as given in the equation below.
\begin{equation}
\begin{split}
    \mathcal{L} \supset \frac{1}{2} \: Y^T m_Y^2 Y \quad  \quad 
    \text{where} \\
    Y \ \equiv  \ \begin{bmatrix}
        \eta_8 \\ \eta_0 \\ \eta_c
    \end{bmatrix} 
    \quad \text{and} \quad 
    m_Y^2  \ = \ \begin{bmatrix}
        m_{88}^2 & m_{08}^2 & 0 \\
        m_{08}^2 & m_{00}^2 & m_{c0}^2 \\
        0        & m_{c0}^2 & m_{cc}^2
    \end{bmatrix} \; .
\end{split}
\label{eq:massmixingETA}
\end{equation}
In \cref{eq:massmixingETA} where we have used semiconventional notation to denote the components of the squared mass matrix, where $0$ and $8$ in the subscripts simply refer to the generators of the coset $U(3)_L\times U(3)_R/U(3)_V$ corresponding to pNGBs $\eta$ and $\eta'$.     

In the limits $m_{08}^2 \ll \left( m_{00}^2 - m_{88}^2\right)$ and $m_{c0}^2 \ll \left( m_{cc}^2 - m_{88}^2\right)$, one can use perturbation theory to find mass eigenvalues and eigenstates. As one expects, only states receive corrections in the first order. Using the hatted field to denote the mass eigenstates, we find
\begin{equation}
\begin{split}
    \hat{Y} \ \equiv  \ \begin{bmatrix}
        \hat{\eta} \\ \hat{\eta}' \\ \hat{\eta}_c
        \end{bmatrix} 
        \ =  \  
        \begin{bmatrix}
         1 - \frac{1}{2}\theta_{08}^2 & \theta_{08} & 0 \\
        - \theta_{08} & 1 - \frac{1}{2}\left( \theta_{08}^2 +\theta_{c0}^2 \right)  
                            & - \theta_{c0} \\
        0        & \theta_{c0} & 1 - \frac{1}{2} \theta_{c0}^2 
    \end{bmatrix} \ Y \;,
\\
\text{where } 
\theta_{08} \ \equiv \frac{m_{08}^2}{m_{00}^2 - m_{88}^2} 
 \ \ll \ 1 \qquad  \text{and} \qquad
\theta_{c0} \ \equiv \ \frac{m_{c0}^2}{m_{cc}^2 - m_{00}^2} \ \ll \ 1 \; .
\end{split} 
\label{eq:statemixingETA}
\end{equation}

We are now in a position to begin the discussion about $B\rightarrow K+Y$ phenomenology, specifically for $B^{+}\rightarrow K^{+}+Y$. The procedure is straightforward. To find the relevant operators one needs only to look at operators which remain nonzero after we take three successive derivatives with respect to $B$, $K$, and $Y$, and then set all fields to zero. We identify four operators that contribute to the transition as described below 
\begin{equation}
   \mathcal{C}_1^6 \: {\Oph_1}^\dag \Oph_6 \ + \ \mathcal{C}_1^7 \: {\Oph_1}^\dag \Oph_7  \ 
   + \ \mathcal{C}_{19} \: \Ophh_{19}  + \ \mathcal{C}_{38} \: \Ophh_{38} 
   \ = \   \sum_a C_a \mathcal{O}^{BKY}_a  \ + \ \cdots  \; ,
\label{eq:BKEtasOperators}
\end{equation}
When the operators of the left hand side of  \cref{eq:BKEtasOperators} are expanded in terms of the component fields one gets a large number of operators. We select a subset of these resultant operators that contribute to $B\rightarrow K+Y$ at the tree level. These operators are denoted here by $\mathcal{O}^{BKY}_a$ where $a$ is a running index. The respective coefficients ($C_a$) are linear combinations of the original Wilson coefficients $\mathcal{C}_1^6, \mathcal{C}_1^7, \mathcal{C}_{19}$, and $\mathcal{C}_{38}$. The rest of the operators (denoted here by $\cdots$) on the right hand side of \cref{eq:BKEtasOperators} play no role.  In \cref{table:BKYoperators} we produce a list of all such operators in $\mathcal{O}^{BKY}_a$ along with their coefficients $C_a$ as defined in \cref{eq:BKEtasOperators}. It is important to point out that in our notation original  Wilson Coefficients (\textit{i.e.}, various $\mathcal{C}$s) are dimensionless (see for \cref{eq:operatorssemileptonic} and \cref{eq:operatorsfullyhadronic}). An extra factor of $4G_F/\sqrt{2}$ per coefficient is required to calculate the amplitudes.   
\begin{table}[ht!]
    \centering
    \begin{tabular}{|c||c|}
    \hline
    Operators $\mathcal{O}^{BKY}_a$ & Coefficients $C_a$ \\
    \hline
$ B^{+} \: \partial_\mu K^{-} \: \partial^\mu \eta_c $  &  
    $- V_{\bar{b}c} V_{\bar{c}s} \  \mathcal{E}_{19} \Lambda_c $ \\
$ \partial_\mu B^{+} \:  K^{-} \: \partial^\mu \eta_c $  &      
    $+ V_{\bar{b}c} V_{\bar{c}s} \  \mathcal{E}_{19} \Lambda_c $ \\
$ \partial_\mu B^{+} \:  \partial^\mu K^{-} \:  \eta_c $  & 0    \\     
    \hline
$ B^{+} \: \partial_\mu K^{-} \: \partial^\mu \eta_0 $  &  
    $ V_{\bar{b}u} V_{\bar{u}s} \ \frac{1}{\sqrt{12}}  
        \left( 2\mathcal{C}_{1}^{7} - \mathcal{E}_{38} \right) f_\pi  $ \\
$ \partial_\mu B^{+} \:  K^{-} \: \partial^\mu \eta_0 $  &
    $ V_{\bar{b}u} V_{\bar{u}s} {\frac{1}{\sqrt{3}}} \ \mathcal{E}_{38} f_\pi 
      \ + \ V_{\bar{b}c} V_{\bar{c}s} \ \frac{2}{\sqrt{3}} 
         \mathcal{E}_{19} \frac{\Lambda_c^2}{f_{\pi}}  $ \\    
 $ \partial_\mu B^{+} \:  \partial^\mu K^{-} \:  \eta_0 $  & 
    $ V_{\bar{b}u} V_{\bar{u}s}\frac{1}{\sqrt{12}} \ 
        \left(- 2 \mathcal{C}_{1}^{7} + \mathcal{E}_{38} \right) f_\pi 
      \ + \ V_{\bar{b}c} V_{\bar{c}s} \ \frac{2}{\sqrt{3}}  
         \mathcal{E}_{19}\ \frac{\Lambda_c^2}{f_{\pi}}  $ \\        
    \hline 
 $ B^{+} \: \partial_\mu K^{-} \: \partial^\mu \eta_8 $  &  
    $ V_{\bar{b}u} V_{\bar{u}s} \ \frac{1}{\sqrt{24}}   
        \left( 2\mathcal{C}_{1}^{7} - \mathcal{E}_{38} \right) f_\pi  $ \\
$ \partial_\mu B^{+} \:  K^{-} \: \partial^\mu \eta_8 $  &
    $ V_{\bar{b}u} V_{\bar{u}s} \ \frac{1}{4\sqrt{6}}   
        \left(\mathcal{E}_{38}-6\ \mathcal{C}_{1}^{7}+12\ \mathcal{C}_{1}^{6}\frac{\Lambda_B}{f_\pi}  \right)f_\pi 
      \ - \ V_{\bar{b}c} V_{\bar{c}s} \ \frac{1}{\sqrt{6}}   
         \mathcal{E}_{19} \frac{\Lambda_c^2}{f_{\pi}}  $ \\    
 $ \partial_\mu B^{+} \:  \partial^\mu K^{-} \:  \eta_8 $  & 0    \\    
    \hline
    \end{tabular}
    \caption{Operators involving $B$, $K$ and $\eta_c,\eta^\prime, \eta$ mesons.}
    \label{table:BKYoperators}
\end{table}    
Starting with the operators in \cref{table:BKYoperators}, calculating various $B^{+}\rightarrow K^{+}Y$ transitions at the leading order is rather straightforward. One only needs to be aware of the transition to a mass basis (or, rather, $Y \rightarrow \hat{Y}$) as given in \cref{eq:statemixingETA}.  In this work, we assume that the angles $\theta_{08}$ and $\theta_{c0}$ are small enough that neglecting terms quadratic in angles is justified. The results are given in \cref{table:BKYamplitudes}.   

\begin{table}[ht]
\renewcommand{\arraystretch}{1.2} % slightly more vertical space
\centering
\begin{tabularx}{\linewidth}{|p{3cm}|X|}
\hline
\textbf{Transitions} & \textbf{Amplitudes} \\
\hline
\[\State{B^+} \to \State{K^+ \eta_c}\] &
%\footnotesize
\[
\frac{4G_F}{\sqrt{2}}\, \mathcal{E}_{19} V_{\bar{b}c} V_{\bar{c}s}\, \Lambda_c\, (m_B^2 - m_K^2)
\]
\normalsize
\\
\hline
\[\State{B^+} \to \State{K^+ \eta'}\] &
%\footnotesize
\[
\begin{aligned}
\frac{4G_F}{\sqrt{2}} \Bigg[ &\mathcal{E}_{19} V_{\bar{b}c} V_{\bar{c}s}\, \Lambda_c \left\{ \theta_{c0} (m_B^2 - m_K^2) + \frac{2}{\sqrt{3}} \frac{\Lambda_c}{f_\pi} m_B^2 \right\} \\
&+ V_{\bar{u}s} V_{\bar{b}u} f_\pi \left\{ \mathcal{E}_{38} \frac{2m_B^2 - m_K^2}{2\sqrt{3}} - \mathcal{C}_{1}^{7} \frac{m_B^2 - m_{\eta'}^2}{\sqrt{3}} \right\}
\Bigg]
\end{aligned}
\]
\normalsize
\\
\hline
\[\State{B^+} \to \State{K^+ \eta}\] &
%\footnotesize
\[
\begin{aligned}
\frac{4G_F}{\sqrt{2}} \Bigg[ 
&-\mathcal{E}_{19} V_{\bar{b}c} V_{\bar{c}s} \frac{\Lambda_c^2}{\sqrt{6}f_\pi} m_B^2 \\
&+ V_{\bar{u}s} V_{\bar{b}u}f_\pi \Bigg\{ 
\mathcal{E}_{38}  \frac{m_B^2 - 2m_K^2}{4\sqrt{6}} 
+ \mathcal{C}_{1}^{6} \frac{\Lambda_B}{f_\pi}\frac{m_\eta^2 - m_K^2} {\sqrt{2/3}}\\
&\quad\quad\quad\quad - \mathcal{C}_{1}^{7}  \frac{2m_B^2 - 3m_K^2 + m_\eta^2}{2\sqrt{6}} 
\Bigg\}
\Bigg]
\end{aligned}
\] 
\normalsize
\\
\hline
\end{tabularx}
\caption{Amplitudes for $|B\rangle \to |K\eta\rangle$ transitions.}
\label{table:BKYamplitudes}
\end{table}

Consider first the rate of $\State{B^{+}} \to \State{K^{+}\eta_c}$. The amplitudes listed in \cref{table:BKpiamplitudes} can be easily converted to decay widths.  Further, using the observed rate, the product of the unknown parameter $\Lambda_c$ and the Wilson coefficient $\mathcal{E}_{19}$ can be estimated. We use the branching width as listed in PDG~\cite{ParticleDataGroup:2024cfk} and find the following.
\begin{equation}
    \Gamma_{B^{+} \to K^{+}\eta_c} \ = \ 4 \times 10^{-16}~\text{GeV} 
    \ \implies \  \mathcal{E}_{19} \Lambda_c \approx 0.012~\text{GeV} \; . 
\end{equation}
We are now ready to tackle the ratios given in \cref{eq:B2KYratios} using the amplitudes listed in \cref{{table:BKYamplitudes}}. Compare first the amplitudes for the decays $B^{+} \to K^{+}\eta/\eta'$. The interesting aspects of the observation related to the hierarchy in these modes can be best understood in the limit $\mathcal{E}_{19} \to 0$ and $\mathcal{E}_{38} \to 0$, when $B^{+} \to K^{+}\eta/\eta'$ proceeds only through the double trace operators (that is, terms proportional to $\mathcal{C}^7_1$ and $\mathcal{C}^6_1$). Then we can immediately argue that $B^{+} \to K^{+}\eta' \gg B^{+} \to K^{+}\eta $ is not generic and can only happen in the event that there are large cancelations among the terms in the amplitude of decay to $\eta$. This, therefore, suggests that single trace operators play crucial roles for decays to $\eta$ and $\eta'$. Now consider an opposite limit where the dominant contribution (at least for decays to $\eta'$) arises from the $ \Ophh_{19}$ operator. We find from \cref{eq:B2KYratios} that    
\begin{equation}
    \theta_{c0}\ \approx  \ \frac{2}{\sqrt{3}} \frac{\Lambda_c}{f_\pi}  \ \approx \ 0.1 \; ,
\end{equation}
where we assume that both terms are of the same order. Remarkably, if we continue to neglect all terms proportional to $V_{\bar{b}u} f_\pi$, we actually end up predicting the ratio of $\Gamma\left( B\rightarrow K+\eta' \right)/ \Gamma\left( B\rightarrow K+\eta \right)$ to be of order $30$ (the same as seen in \cref{eq:B2KYratios}). Generically speaking, we expect to produce both ratios in \cref{eq:B2KYratios} as long as the single trace operators dominate in the decay to $\eta'$, and the terms proportional to $V_{\bar{b}u} f_\pi$ are comparable or subdominant than the term proportional to $\mathcal{E}_{19}$. 

Summarizing our results:
\begin{enumerate}
    \item The decay of $B\to K+\eta_c$ proceeds via a single trace operator (namely, $\Ophh_{19}$) and the rate is proportional to a scale from the charm sector (namely, $\Lambda_c$). The observed rate of $B\to K+\eta_c$ allows us to estimate the product of $\Lambda_c$ and a Wilson coefficient (namely, $\mathcal{E}_{19} \Lambda_c \approx 0.012~\text{GeV}$).  
    \item The $B\to K+\eta/\eta'$ system is under-constrained in our framework, given that ultimately $4$ operators contribute to the full amplitude. However, several conclusions can be drawn if we require that there is no large cancellation that ultimately yields the hierarchies in \cref{eq:B2KYratios} (let us call it \emph{natural ansatz}). For example, if $B\to K+\eta'$ becomes saturated by one of the terms proportional to $\mathcal{E}_{19}$, one gets estimations of the sizes of $\theta_{c0}$ or $\Lambda_c$.    
    \begin{equation}
        \theta_{c0}  \ \lesssim \ 0.2 \; , \quad  \Lambda_c \ \lesssim \ 0.2 \: f_\pi \; , \quad \text{and therefore,} \quad 
        \mathcal{E}_{19} \  \gtrsim 0.5  \; .
    \end{equation} 
    \item Building on similar logic, we also find that $\Ophh_{38}$ can only contribute if $\mathcal{E}_{38} \sim \Order{1}$ because of the relative $\lambda^2$ additional suppression of CKM elements, where $\lambda$ refer to the Wolfenstein parameter. 
    \item The contribution of the double trace operator ${\Oph_1}^\dag \Oph_7$ to $B\to K+\eta'$ must be suppressed with respect to other terms (without a big cancelation ansatz), otherwise $B\to K+\eta$ becomes too large. This suggests $\mathcal{C}^7_1 \ll 1$.  
    \item Finally, the contribution of the double trace operator ${\Oph_1}^\dag \Oph_6$ to $B\to K+\eta$ is additionally suppressed by $m_\eta^2/m_B^2$ with respect to other terms (apart from suppression of the CKM elements). Within the realm of natural ansatz, we do not see it playing an important role in these decays.   
    
\end{enumerate}

Before concluding, let us mention several curious observations. 
\begin{enumerate}
    \item The high rate of $B\to K+\eta_c$ and the imposition of the natural ansatz suggest that the Wilson coefficient of the single trace operator $\Ophh_{19}$ is rather large (i.e. $\mathcal{E}_{19} \  \sim \Order{0.5}$). Given that  $\Ophh_{19}$  is a color suppressed operator, it is curious. 
    \item Even if we take $\theta_{c0} \to 0$ -- that is, without mixing of $\eta'$ and $\eta_c$, we still find that the same operator that generates $B\to K+\eta_c$ generates $B\to K+\eta'$ as well.     
    \item{$B^0 D_S^{+}K^-$ process: }
\begin{equation}
    \mathcal{M}= -i \frac{4G_F}{\sqrt{2}}V_{\bar{c}b}V_{\bar{d}u}\Bigg[\Lambda_b \mathcal{E}_{22}\; \underbrace{p_D.\left(p_B+p_K\right)}_{m_B^2-m_K^2}+f_\pi \mathcal{E}_{23}\; \underbrace{\frac{p_K\left(p_B+p_D\right)}{2}}_{\frac{m_B^2-m_D^2}{2}}\Bigg]\;.
\end{equation}
This mode, in the diagrammatic picture of the quarks, proceeds via what is called the $W$-exchange and is expected to be quite suppressed (of order $\sim f_B/m_B$). Consequently, it makes this mode a good place to look for rescattering effects \cite{Blok:1997yj}, which proceeds via first the decay $B^0\to D^-\pi^+$ and then via the rescattering of the $D\pi$ final state, (with $K,\,K^*$ exchange, for example) to $D_s^{-}K^+$ or non-factorizable contributions \cite{Mahajan:2005pc}. In the amplitude above, the first term, proportional to $\Lambda_b$ is expected to be small (recall that $\Lambda_c \sim 0.1f_{\pi}$, and we expect $\Lambda_b$ to be even smaller). In addition, $m_B^2-m_K^2 \sim 27$ is not much larger than $m_B^2-m_{D_s}^2 \sim 24$ . Approximating the amplitude with the second term and comparing it with the observed branching ratio leads to $\mathcal{E}_{23} \sim 0.02$. 
\end{enumerate}

%------------------------------------------------------
\section{Conclusions and future directions}
\label{sec:conclusion}
%------------------------------------------------------

In this work, we have developed a systematic phenomenological Lagrangian framework for charged–current weak decays of heavy–light pseudoscalar mesons, combining the approximate chiral symmetry of the light sector with heavy–quark flavor symmetry. Treating the elements of the CKM matrix as spurions, we constructed the most general set of leading-order current–current operators at $\mathcal{O}(G_F)$, organized according to their transformation properties in the approximate global symmetry group. Our analysis was restricted to charged–current interactions and pseudoscalar mesons, allowing us to exhaustively classify the operator basis at dimension six.

We showed that leptonic and semileptonic decays of heavy–light mesons are described by a small and well-defined subset of operators, characterized by a limited number of Wilson coefficients. By matching onto known results from HQET and heavy-quark spin symmetry, we demonstrated that our framework reproduces the correct scaling behavior of decay constants and form factors in the appropriate kinematic limits. In particular, the framework naturally accommodates the heavy-quark mass dependence of $B$, $D$, and $B_c$ leptonic and semileptonic decays, providing non-trivial consistency checks on the construction.

For nonleptonic decays, we identified a rich operator structure already at leading order, consisting of both double-trace and single-trace operators. We demonstrated explicitly that the resulting amplitudes satisfy the well-known isospin sum rules for $B \to K\pi$ and $B \to D\pi$ transitions, confirming that the effective Lagrangian correctly encodes the underlying flavor symmetries. The mapping between operator coefficients and topological amplitudes further clarifies how tree-level, color-suppressed, exchange, and penguin-like contributions arise within a unified phenomenological description. An interesting application of our framework is the analysis of $B\to K+Y$, where $Y=\eta,\eta^\prime,\eta_c$. Treating these states as flavor eigenstates with perturbative mass mixing, we demonstrate that single-trace operators play a crucial role in explaining the observed hierarchy of branching ratios. The same operator responsible for the sizable $B\to K \eta_c$ rate also contributes to other $\eta$ modes and therefore provides an explanation of the hierarchies of the observed branching fractions. The resulting constraints on the mixing angles and Wilson coefficients are derived from the experimental data and from the minimal naturalness assumption. 

Overall, our work demonstrates that a phenomenological Lagrangian built directly from the mesonic degrees of freedom and guided by approximate symmetries provides a powerful and complementary way of studying various phenomenologically interesting processes and quark level factorization approaches. Although the framework is limited to charged current processes and pseudoscalar mesons, it is readily extendable. Natural extensions include the incorporation of neutral current operators, vector mesons, etc. Such extensions will broaden the phenomenological reach of the framework and are being pursued elsewhere~\cite{chakraborty:future}.

%------------------------------------------------------
\acknowledgments
The authors thank Siddhartha Karmakar for helpful discussions. NM and TSR thank the organizers of the workshop Heavy Hadrons in Heavy Ion and Particle Collisions (HHHPCS 2023) at IIT Gandhinagar where discussions along these lines were initiated. The work by NM at the Physical Research Laboratory is supported by the Department of Space (DoS), Government of India. NM also acknowledges the partial support under the MATRICS project (MTR/023/000442) from the Science $\&$ Engineering Research Board (SERB), Department of Science and Technology (DST), Government of India.
SC thanks the IIT-Kanpur initiation grant (PHY/2022220) and the Science and Engineering Research Board, Government of India (Grant No. SRG/2023/001162) for financial support. 

%------------------------------------------------------
\appendix
\section{The set-up}
\label{sec:app1}
%------------------------------------------------------

To describe electroweak interactions of quarks in a manifestly $G_{\text{approx}} \ \equiv \ U(3)_L \times U(3)_R \times SU(2)_H$ symmetric way, we treat light quarks ($u,d,s$) and heavy quarks ($c,b$) separately. In particular, consider Weyl spinors (all left-handed) representing SM quarks written in the mass basis and their transformations under $G_{\text{approx}} $. 
\begin{equation}
\begin{split}
    q \equiv \begin{pmatrix} u \\ d \\ s \end{pmatrix}\, ,  \quad 
    q^c \equiv \begin{pmatrix} u^c & d^c & s^c \end{pmatrix}  
        \qquad \text{and} \qquad
    Q_H \equiv \begin{pmatrix} c \\ b  \end{pmatrix}\, ,  \quad
    Q^c_H \equiv \begin{pmatrix} c^c & b^c  \end{pmatrix}  \\
    \text{where}  \qquad 
    q \rightarrow L q \, , \qquad 
    q^c \rightarrow q^c R^\dag \, , \qquad
    Q_H \rightarrow V_H Q_H \, , \qquad 
    Q^c_H \rightarrow Q^c_H V_H^\dag \, , \qquad
\end{split}  
\label{eq:defQfields}   
\end{equation}

Using the electroweak charge matrices introduced in \eqref{eq:defWcharges} we can write the quark interactions of $W_{\pm}$ in a suggestive way 
\begin{equation}
    \left( q^\dag  \hat{Q}_{qq}  \bar{\sigma}_\mu q  \ + \ 
    Q^\dag  \hat{Q}_{QQ} \bar{\sigma}^\mu Q  \ + \ 
    q^\dag  \hat{Q}_{qQ} \bar{\sigma}^\mu Q  \ + \  
    Q^\dag  \hat{Q}_{Qq} \bar{\sigma}^\mu q  \right) \frac{g_2}{\sqrt{2}} W_{-}^\mu  
    \ + \ \text{h.c.} \; ,
\label{eq:intW}    
\end{equation}
where $g_2$ is the gauge coupling constant, $\hat{Q}_{qq}, \hat{Q}_{qQ}, \hat{Q}_{Qq}$, and $\hat{Q}_{QQ}$ are previously defined charge matrices, whose nonzero components involve various CKM matrix elements. For completeness, note that the interactions of the $W_{+}$ field are in the Hermitian conjugate pieces. 

If we treat the charge matrices as spurions, with the rule of transformation given in \eqref{eq:defWchargestransformation}, it is straightforward to check that \eqref{eq:intW} is manifestly invariant under $G_{\text{approx}}$ transformations. 

\subsection{Semi-leptonic Operators} 

In energies below the electroweak scale, the EFT (with the massive electroweak bosons) at the tree level simply follows as one replaces $W_{\pm}$ by the corresponding equation of motion (EOM). In other words, the effective operators simply follow from a current-current interaction when the corresponding electroweak currents (namely, $J_{W_\pm}$) are expanded in terms of light quarks and leptons. For example, the semileptonic operators involving quarks and leptons simply arise from
\begin{equation}
    \frac{G_F}{\sqrt{2}} \: J_{W_{+}} J_{W_{-}} \supset  \frac{G_F}{\sqrt{2}} \: \left( q^\dag  \hat{Q}_{qq}  \bar{\sigma}_\mu q  + 
    Q^\dag  \hat{Q}_{QQ} \bar{\sigma}^\mu Q   +  
    q^\dag  \hat{Q}_{qQ} \bar{\sigma}^\mu Q   +   
    Q^\dag  \hat{Q}_{Qq} \bar{\sigma}^\mu q  \right) \left( \nu^\dag \bar{\sigma}^\mu \ell \right) + \dots
\label{eq:quarksemileptonicOperators}
\end{equation}
The challenge is to convert these quark-lepton operators into hadron-lepton ones. As far as light-light mesons (or the pion-nonets) are considered, it is well understood and the answer lies in the formation of the \ChPT Lagrangian. In the following, we work it out using our degree of freedom (namely, $\Sigma$). Even though these steps are textbook materials, we repeat these since our nontrivial constructions emerge from these well-traversed paths. 

Rewriting $\hat{Q}_{qq} = \sum_a \hat{Q}_{qq}^a t^a $, $t^a$ being generators of $U(3)_L$ one can write the quark in a suggestive pattern.
\begin{equation}
    q^\dag  \hat{Q}_{qq}  \bar{\sigma}_\mu q = \sum_a \left( q^\dag  t^a  \bar{\sigma}_\mu q \right) \hat{Q}_{qq}^a = \sum_a J_{L_\mu}^a \hat{Q}_{qq}^a \; , 
\end{equation}
where $J_L^a$ is the current for the global $U(3)_L$. The trick to derive the low energy interaction of hadrons (at low energy) is to simply match the quark current for a given symmetry generator with the corresponding hadron current. It is, therefore, straightforward to derive the semi-leptonic operator for light-light mesons at the lowest order after noting the transformation property of the $\Sigma$ field.
\begin{equation}
\begin{split}
   q^\dag  \hat{Q}_{qq}  \bar{\sigma}_\mu q  
                \left( \nu^\dag \bar{\sigma}^\mu \ell \right)   
    \ = \ \sum_a J_{L_\mu}^a \hat{Q}_{qq}^a  
                \left( \nu^\dag \bar{\sigma}^\mu \ell \right)   \\
    \ = \ \frac{\iu}{2} \tr{ \Sigma^\dag t^a \overleftrightarrow{\partial}_\mu \Sigma } 
                \hat{Q}_{qq}^a  \left( \nu^\dag \bar{\sigma}^\mu \ell \right) \ + \ \cdots 
                \\ 
    \ = \ \frac{\iu}{2} \tr{ \Sigma^\dag \hat{Q}_{qq} 
            \overleftrightarrow{\partial}_\mu \Sigma } 
             \left( \nu^\dag \bar{\sigma}^\mu \ell \right) \ + \ \cdots 
                           \; ,   
\end{split}
\label{eq:Qqqsemileptonic1}
\end{equation}
where $\cdots$ represents additional terms. In \ChPT, one achieves the same operator from $\Order{p^4}$ by appropriately gauging the global $SU(3)_L$ and consequently replacing the partial derivative with covariant ones. In fact, demanding that we regain the exact Fermi suppressed semi-leptonic operators of \ChPT gives the relationship between the $\Sigma$ field and the non-linear $U_\pi$ as stated in \cref{eq:defCompFieldsSigma}. 

However, note that the expression in \cref{eq:Qqqsemileptonic1} is still incomplete. According to the discussion before \cref{eq:defCompFieldsH}, the field $\Phi$ also furnishes the fundamental representation of $U(3)_L$. Consequently, one additionally expects a current with the field $\Phi$, which, by construction, gives an operator with $H$ because of the replacement rule $\Phi \to H^\dag$. 
\begin{equation}
   q^\dag  \hat{Q}_{qq}  \bar{\sigma}_\mu q   
    \ = \ \sum_a J_{L_\mu}^a \hat{Q}_{qq}^a                     
    \ \supset \ 
    \frac{\iu}{2} \tr{ \Phi^\dag \hat{Q}_{qq} \overleftrightarrow{\partial}_\mu \Phi }
    \ \propto \ 
     \tr{ H^\dag \hat{Q}_{qq} \overleftrightarrow{\partial}_\mu H }   \; ,   
\label{eq:Qqqsemileptonic2}
\end{equation}
This gives $\Oph_{1}$ and $\Oph_{4}$ in \cref{table:hadroniccurrents}. 

In \ChPT, one uses the current matching not just to identify the operator but also to determine the strength of the operator (i.e., the corresponding Wilson Coefficient). As mentioned many times in the main text of this work, our aim is only to identify operators. In this we realize that one could have come to the operator identification simply by searching for an operator of mass dimension $3$ which contains the charge matrix $\hat{Q}_{qq}$ and a single factor of  $\overleftrightarrow{\partial}_{\!\!\!\mu}$ and is invariant under $G_{\text{approx}}$. We take this rather simpler approach to construct rest of the operators in \cref{table:hadroniccurrents}.   

For the sake of completeness let us also chalk out the procedure for identifying currents via the symmetry route. For this we will construct groups of symmetry larger than the $G_{\text{approx}}$. First start with extending the vectorial $SU(2)_H$ to $SU(2)_{L_H} \times SU(2)_{R_H}$. The transformation properties of $\Sigma_H$, $\Phi$ and $\Phi_c$ needs to be defined again
\begin{equation}
    \Sigma_H \rightarrow  L_H  \Sigma_H R_H^\dag  \; , \quad 
    \Phi  \rightarrow  L  \Phi R_H^\dag \; , \quad \text{and} \quad
    \Phi_c \rightarrow  L_H  \Phi R^\dag \; ,
\end{equation}
with $L_H$ and $R_H$ representing elements of $SU(2)_{L_H}$ and $ SU(2)_{R_H}$ respectively. For the electroweak charge matrices, the new set of transformation laws can be noted by simply replacing $V_H \to L_H$. It is straightforward to replicate the same procedure as shown in \crefrange{eq:Qqqsemileptonic1}{eq:Qqqsemileptonic2} after utilizing currents for $SU(2)_{L_H}$. Noting that both $\Sigma_H$ and $\Phi_c$ transform according to the fundamentals of $U(2)_{L_H}$ we can guess the form of hadronic semiloptonic operators as long as the matching between the quark currents and the hadronic ones works. In particular, consider the quark operators in \cref{eq:quarksemileptonicOperators}  proportional to the charge matrix $\hat{Q}_{QQ}$. Noting that this charge matrix can be expanded in generators of $SU(2)_{L_H}$ we can derive the corresponding operators. 
\begin{equation}
\begin{split}   
   Q^\dag  \hat{Q}_{QQ}  \bar{\sigma}_\mu Q        
    \ = \ \sum_a J_{L_{H_\mu}}^a \hat{Q}_{QQ}^a 
    \ = \ \frac{\iu}{2} \Big[  \tr{ \Sigma_H^\dag \hat{Q}_{QQ} \overleftrightarrow{\partial}_\mu \Sigma_H
     +    \Phi_c^\dag \hat{Q}_{QQ} \overleftrightarrow{\partial}_\mu \Phi_c } 
         \Big]     
    \ + \ \cdots
    \; ,   
\end{split}
\label{eq:QQQsemileptonic}
\end{equation}
where $t^a$ are generators of the global $SU(2)_{H_L}$ of heavy quarks, and $J_{L_{H_\mu}}^a$ are the corresponding currents. Upon replacing $\Phi_c \to H$ this gives $\Oph_{2}$ and $\Oph_{3}$ in \cref{table:hadroniccurrents}.

A far trickier problem is to come up with operators proportional to charge matrices $\hat{Q}_{qQ}$ and $\hat{Q}_{Qq}$.  In order to have an understanding of the form of operators one might expect, we embed $SU(3)_L \times SU(3)_R \times SU(2)_{L_H} \times SU(2)_{R_H}$ into $SU(5)_L\times SU(5)_R$, put $q$ and $Q_H$ into a fundamental of $SU(5)_L$, and put $q^c$ and $Q_H^c$ into an anti-fundamental representation of $SU(5)_R$. An heavy-light current then corresponds to a flavor current for one of the generators in the coset $SU(5)_L /\left( SU(3)_L \times SU(2)_{L_H} \right)$. Embedding meson fields ($\Sigma, \Sigma_H, \Phi$ and $\Phi_c$) into a single $\left( 5, \bar{5}\right)$ of $SU(5)_L\times SU(5)_R$, we find the corresponding currents given in terms of the meson fields. Once again replacing  $\Phi_c \to H$ and $\Phi \to H^\dag$ gives $\Oph_{5,6,7,8}$ in \cref{table:hadroniccurrents}.

\bibliographystyle{JHEP}
\bibliography{phenoflavor.bib}

@article{chakraborty:future,
    author = "Chakraborty, Sabyasachi and Karmakar, Siddhartha and Mahajan, Namit and Roy, Tuhin S.",
    title = "{TBD}"
}

@article{Csaki:future,
    author = "Cs{\'a}ki, Csaba and Roy, Tuhin S. and Ruhdorfer, Maximilian and Youn, Taewook",
    title = "{TBD}"
}

@article{Lucha:1991vn,
    author = "Lucha, W. and Schoberl, F. F. and Gromes, D.",
    title = "{Bound states of quarks}",
    doi = "10.1016/0370-1573(91)90001-3",
    journal = "Phys. Rept.",
    volume = "200",
    pages = "127--240",
    year = "1991"
}

@article{ParticleDataGroup:2024cfk,
    author = "Navas, S. and others",
    collaboration = "Particle Data Group",
    title = "{Review of particle physics}",
    doi = "10.1103/PhysRevD.110.030001",
    journal = "Phys. Rev. D",
    volume = "110",
    number = "3",
    pages = "030001",
    year = "2024"
}

@article{Isgur:1990yhj,
    author = "Isgur, Nathan and Wise, Mark B.",
    title = "{Weak transition form factors between heavy mesons}",
    reportNumber = "UTPT-90-01, CALT-68-1608",
    doi = "10.1016/0370-2693(90)91219-2",
    journal = "Phys. Lett. B",
    volume = "237",
    pages = "527--530",
    year = "1990"
}

@article{Bali:2000gf,
    author = "Bali, Gunnar S.",
    title = "{QCD forces and heavy quark bound states}",
    eprint = "hep-ph/0001312",
    archivePrefix = "arXiv",
    reportNumber = "HUB-EP-99-67",
    doi = "10.1016/S0370-1573(00)00079-X",
    journal = "Phys. Rept.",
    volume = "343",
    pages = "1--136",
    year = "2001"
}

@article{Alkofer:2006fu,
    author = "Alkofer, Reinhard and Greensite, J.",
    title = "{Quark Confinement: The Hard Problem of Hadron Physics}",
    eprint = "hep-ph/0610365",
    archivePrefix = "arXiv",
    doi = "10.1088/0954-3899/34/7/S02",
    journal = "J. Phys. G",
    volume = "34",
    pages = "S3",
    year = "2007"
}

@article{Brambilla:2014jmp,
    author = "Brambilla, N. and others",
    title = "{QCD and Strongly Coupled Gauge Theories: Challenges and Perspectives}",
    eprint = "1404.3723",
    archivePrefix = "arXiv",
    primaryClass = "hep-ph",
    reportNumber = "CCQCN-2014-24, CCTP-2014-5, CERN-PH-TH-2014-033, DF-1-2014, HIP-2014-03-TH, ITEP-LAT-2014-1, JLAB-THY-14-1865, MITP-14-016, NT@UW-14-04, RUB-TPII-01-2014, TUM-EFT-46-14, FERMILAB-PUB-14-024-T, LLNL-JRNL-651216, UWTHPH-2014-006",
    doi = "10.1140/epjc/s10052-014-2981-5",
    journal = "Eur. Phys. J. C",
    volume = "74",
    number = "10",
    pages = "2981",
    year = "2014"
}

@article{Mahajan:2005pc,
    author = "Mahajan, Namit",
    title = "{W-exchange/annihilation amplitudes in LCSR: B(d)0 ---{\ensuremath{>}} D(s)- K+ as an example}",
    eprint = "hep-ph/0508230",
    archivePrefix = "arXiv",
    doi = "10.1016/j.physletb.2006.01.051",
    journal = "Phys. Lett. B",
    volume = "634",
    pages = "240--248",
    year = "2006"
}

@article{Blok:1997yj,
    author = "Blok, Boris and Gronau, Michael and Rosner, Jonathan L.",
    title = "{Annihilation, rescattering, and CP asymmetries in B meson decays}",
    eprint = "hep-ph/9701396",
    archivePrefix = "arXiv",
    reportNumber = "EFI-97-04, TECHNION-PH-97-3, TECHNION-PH-97-03",
    doi = "10.1103/PhysRevLett.78.3999",
    journal = "Phys. Rev. Lett.",
    volume = "78",
    pages = "3999--4002",
    year = "1997"
}

@article{Lepage:1992tx,
    author = "Lepage, G. Peter and Magnea, Lorenzo and Nakhleh, Charles and Magnea, Ulrika and Hornbostel, Kent",
    title = "{Improved nonrelativistic QCD for heavy quark physics}",
    eprint = "hep-lat/9205007",
    archivePrefix = "arXiv",
    reportNumber = "CLNS-92-1136, OHSTPY-HEP-T-92-001",
    doi = "10.1103/PhysRevD.46.4052",
    journal = "Phys. Rev. D",
    volume = "46",
    pages = "4052--4067",
    year = "1992"
}

@article{FlavourLatticeAveragingGroupFLAG:2024oxs,
    author = "Aoki, Y. and others",
    collaboration = "Flavour Lattice Averaging Group (FLAG)",
    title = "{FLAG review 2024}",
    eprint = "2411.04268",
    archivePrefix = "arXiv",
    primaryClass = "hep-lat",
    reportNumber = "CERN-TH-2024-192, FERMILAB-PUB-24-0785-T",
    doi = "10.1103/nfzp-p5dn",
    journal = "Phys. Rev. D",
    volume = "113",
    number = "1",
    pages = "014508",
    year = "2026"
}

@article{Aoki:2016frl,
    author = "Aoki, S. and others",
    title = "{Review of lattice results concerning low-energy particle physics}",
    eprint = "1607.00299",
    archivePrefix = "arXiv",
    primaryClass = "hep-lat",
    reportNumber = "CP3-Origins-2016-023, DESY-16-111, DIAS-2016-23, Edinburgh-2016-11, FTUAM-16-23, HIM-2016-02, IFT-UAM-CSIC-16-057, LPT-Orsay-16-47, MITP-16-059, RM3-TH-16-7, ROM2F-2016-05, YITP-16-77",
    doi = "10.1140/epjc/s10052-016-4509-7",
    journal = "Eur. Phys. J. C",
    volume = "77",
    number = "2",
    pages = "112",
    year = "2017"
}

@article{Tsang:2023nay,
    author = "Tsang, J. Tobias and Della Morte, Michele",
    title = "{B-physics from lattice gauge theory}",
    eprint = "2310.02705",
    archivePrefix = "arXiv",
    primaryClass = "hep-lat",
    reportNumber = "CERN-TH-2023-175",
    doi = "10.1140/epjs/s11734-023-01011-3",
    journal = "Eur. Phys. J. ST",
    volume = "233",
    number = "2",
    pages = "253--270",
    year = "2024"
}

@article{USQCD:2022mmc,
    author = "Kronfeld, Andreas S. and others",
    collaboration = "USQCD",
    title = "{Lattice QCD and Particle Physics}",
    eprint = "2207.07641",
    archivePrefix = "arXiv",
    primaryClass = "hep-lat",
    reportNumber = "FERMILAB-CONF-22-531-T",
    month = "7",
    year = "2022"
}

@article{Pich:1995bw,
    author = "Pich, A.",
    title = "{Chiral perturbation theory}",
    eprint = "hep-ph/9502366",
    archivePrefix = "arXiv",
    reportNumber = "FTUV-95-4, IFIC-95-4",
    doi = "10.1088/0034-4885/58/6/001",
    journal = "Rept. Prog. Phys.",
    volume = "58",
    pages = "563--610",
    year = "1995"
}

@article{Ecker:1994gg,
    author = "Ecker, G.",
    title = "{Chiral perturbation theory}",
    eprint = "hep-ph/9501357",
    archivePrefix = "arXiv",
    reportNumber = "UWTHPH-1994-49",
    doi = "10.1016/0146-6410(95)00041-G",
    journal = "Prog. Part. Nucl. Phys.",
    volume = "35",
    pages = "1--80",
    year = "1995"
}

@article{Scherer:2002tk,
    author = "Scherer, Stefan",
    editor = "Negele, John W. and Vogt, E. W.",
    title = "{Introduction to chiral perturbation theory}",
    eprint = "hep-ph/0210398",
    archivePrefix = "arXiv",
    reportNumber = "MKPH-T-02-09",
    journal = "Adv. Nucl. Phys.",
    volume = "27",
    pages = "277",
    year = "2003"
}

@article{Politzer:1988bs,
    author = "Politzer, H. David and Wise, Mark B.",
    title = "{Effective Field Theory Approach to Processes Involving Both Light and Heavy Fields}",
    reportNumber = "CALT-68-1494",
    doi = "10.1016/0370-2693(88)90656-9",
    journal = "Phys. Lett. B",
    volume = "208",
    pages = "504--507",
    year = "1988"
}

@article{Eichten:1989zv,
    author = "Eichten, Estia and Hill, Brian Russell",
    title = "{An Effective Field Theory for the Calculation of Matrix Elements Involving Heavy Quarks}",
    reportNumber = "FERMILAB-PUB-89-184-T",
    doi = "10.1016/0370-2693(90)92049-O",
    journal = "Phys. Lett. B",
    volume = "234",
    pages = "511--516",
    year = "1990"
}

@article{Georgi:1990um,
    author = "Georgi, Howard",
    title = "{An Effective Field Theory for Heavy Quarks at Low-energies}",
    reportNumber = "HUTP-90/A007",
    doi = "10.1016/0370-2693(90)91128-X",
    journal = "Phys. Lett. B",
    volume = "240",
    pages = "447--450",
    year = "1990"
}

@article{Grinstein:1990mj,
    author = "Grinstein, Benjamin",
    title = "{The Static Quark Effective Theory}",
    reportNumber = "HUTP-90/A002",
    doi = "10.1016/0550-3213(90)90349-I",
    journal = "Nucl. Phys. B",
    volume = "339",
    pages = "253--268",
    year = "1990"
}

@article{Neubert:1993mb,
    author = "Neubert, Matthias",
    title = "{Heavy quark symmetry}",
    eprint = "hep-ph/9306320",
    archivePrefix = "arXiv",
    reportNumber = "SLAC-PUB-6263",
    doi = "10.1016/0370-1573(94)90091-4",
    journal = "Phys. Rept.",
    volume = "245",
    pages = "259--396",
    year = "1994"
}

@article{Bigi:1997fj,
    author = "Bigi, Ikaros I. Y. and Shifman, Mikhail A. and Uraltsev, N.",
    title = "{Aspects of heavy quark theory}",
    eprint = "hep-ph/9703290",
    archivePrefix = "arXiv",
    reportNumber = "TPI-MINN-97-02-T, UMN-TH-1528-97, UND-HEP-97-BIG01",
    doi = "10.1146/annurev.nucl.47.1.591",
    journal = "Ann. Rev. Nucl. Part. Sci.",
    volume = "47",
    pages = "591--661",
    year = "1997"
}

@book{Manohar:2000dt,
    author = "Manohar, Aneesh V. and Wise, Mark B.",
    title = "{Heavy quark physics}",
    doi = "10.1017/9781009402125",
    isbn = "978-0-521-03757-0, 978-1-009-40212-5",
    volume = "10",
    year = "2000"
}

@article{Mannel:2004ce,
    author = "Mannel, T.",
    title = "{Effective Field Theories in Flavor Physics}",
    doi = "10.1007/b62268",
    journal = "Springer Tracts Mod. Phys.",
    volume = "203",
    pages = "1--175",
    year = "2004"
}

@article{Buchalla:1995vs,
    author = "Buchalla, Gerhard and Buras, Andrzej J. and Lautenbacher, Markus E.",
    title = "{Weak Decays beyond Leading Logarithms}",
    eprint = "hep-ph/9512380",
    archivePrefix = "arXiv",
    reportNumber = "SLAC-PUB-7009, SLAC-PUB-95-7009, MPI-PH-95-104, TUM-T31-100-95, FERMILAB-PUB-95-305-T",
    doi = "10.1103/RevModPhys.68.1125",
    journal = "Rev. Mod. Phys.",
    volume = "68",
    pages = "1125--1144",
    year = "1996"
}

@inproceedings{Buras:1998raa,
    author = "Buras, Andrzej J.",
    title = "{Weak Hamiltonian, CP violation and rare decays}",
    booktitle = "{Les Houches Summer School in Theoretical Physics, Session 68: Probing the Standard Model of Particle Interactions}",
    eprint = "hep-ph/9806471",
    archivePrefix = "arXiv",
    reportNumber = "TUM-HEP-316-98",
    pages = "281--539",
    month = "6",
    year = "1998"
}

@article{Blok:1992na,
    author = "Blok, B. and Shifman, Mikhail A.",
    title = "{Nonfactorizable amplitudes in weak nonleptonic decays of heavy mesons}",
    eprint = "hep-ph/9205221",
    archivePrefix = "arXiv",
    reportNumber = "NSF-ITP-92-76, TPI-MINN-92-22-T",
    doi = "10.1016/0550-3213(93)90330-R",
    journal = "Nucl. Phys. B",
    volume = "389",
    pages = "534--548",
    year = "1993"
}

@article{Neubert:1997uc,
    author = "Neubert, Matthias and Stech, Berthold",
    editor = "Buras, A. J. and Lindner, M.",
    title = "{Nonleptonic weak decays of B mesons}",
    eprint = "hep-ph/9705292",
    archivePrefix = "arXiv",
    reportNumber = "CERN-TH-97-099, CERN-TH-97-99, HD-THEP-97-23",
    doi = "10.1142/9789812812667_0004",
    journal = "Adv. Ser. Direct. High Energy Phys.",
    volume = "15",
    pages = "294--344",
    year = "1998"
}

@article{Cheng:1998uy,
    author = "Cheng, Hai-Yang and Tseng, B.",
    title = "{Nonfactorizable effects in spectator and penguin amplitudes of hadronic charmless B decays}",
    eprint = "hep-ph/9803457",
    archivePrefix = "arXiv",
    reportNumber = "IP-ASTP-01-98",
    doi = "10.1103/PhysRevD.58.094005",
    journal = "Phys. Rev. D",
    volume = "58",
    pages = "094005",
    year = "1998"
}

@article{Ali:1998eb,
    author = "Ali, Ahmed and Kramer, G. and Lu, Cai-Dian",
    title = "{Experimental tests of factorization in charmless nonleptonic two-body B decays}",
    eprint = "hep-ph/9804363",
    archivePrefix = "arXiv",
    reportNumber = "DESY-98-041",
    doi = "10.1103/PhysRevD.58.094009",
    journal = "Phys. Rev. D",
    volume = "58",
    pages = "094009",
    year = "1998"
}

@article{Isgur:1988gb,
    author = "Isgur, Nathan and Scora, Daryl and Grinstein, Benjamin and Wise, Mark B.",
    title = "{Semileptonic B and D Decays in the Quark Model}",
    reportNumber = "UTPT-88-12",
    doi = "10.1103/PhysRevD.39.799",
    journal = "Phys. Rev. D",
    volume = "39",
    pages = "799--818",
    year = "1989"
}

@article{Isgur:1989vq,
    author = "Isgur, Nathan and Wise, Mark B.",
    title = "{Weak Decays of Heavy Mesons in the Static Quark Approximation}",
    reportNumber = "UTPT-89-27, CALT-68-1585",
    doi = "10.1016/0370-2693(89)90566-2",
    journal = "Phys. Lett. B",
    volume = "232",
    pages = "113--117",
    year = "1989"
}

@article{Isgur:1990kf,
    author = "Isgur, Nathan and Wise, Mark B.",
    title = "{Relationship Between Form-factors in Semileptonic $\bar{B}$ and $D$ Decays and Exclusive Rare $\bar{B}$ Meson Decays}",
    reportNumber = "CALT-68-1625, UTPT-90-02",
    doi = "10.1103/PhysRevD.42.2388",
    journal = "Phys. Rev. D",
    volume = "42",
    pages = "2388--2391",
    year = "1990"
}

@article{Chernyak:1983ej,
    author = "Chernyak, V. L. and Zhitnitsky, A. R.",
    title = "{Asymptotic Behavior of Exclusive Processes in QCD}",
    reportNumber = "IYF-83-103, IYF-83-104, IYF-83-105, IYF-83-106, IYF-83-107, IYF-83-108",
    doi = "10.1016/0370-1573(84)90126-1",
    journal = "Phys. Rept.",
    volume = "112",
    pages = "173",
    year = "1984"
}

@article{Wirbel:1985ji,
    author = "Wirbel, M. and Stech, B. and Bauer, Manfred",
    title = "{Exclusive Semileptonic Decays of Heavy Mesons}",
    reportNumber = "HD-THEP-85-19",
    doi = "10.1007/BF01560299",
    journal = "Z. Phys. C",
    volume = "29",
    pages = "637",
    year = "1985"
}

@article{Bauer:1986bm,
    author = "Bauer, Manfred and Stech, B. and Wirbel, M.",
    title = "{Exclusive Nonleptonic Decays of D, D(s), and B Mesons}",
    reportNumber = "HD-THEP-86-19, DO-THEP-86-21",
    doi = "10.1007/BF01561122",
    journal = "Z. Phys. C",
    volume = "34",
    pages = "103",
    year = "1987"
}

@article{Stoler:1993yk,
    author = "Stoler, P.",
    title = "{Baryon form-factors at high Q**2 and the transition to perturbative QCD}",
    doi = "10.1016/0370-1573(93)90088-U",
    journal = "Phys. Rept.",
    volume = "226",
    pages = "103--171",
    year = "1993"
}

@article{Falk:1990yz,
    author = "Falk, Adam F. and Georgi, Howard and Grinstein, Benjamin and Wise, Mark B.",
    title = "{Heavy Meson Form-factors From {QCD}}",
    reportNumber = "HUTP-90/A011, CALT-68-1618",
    doi = "10.1016/0550-3213(90)90591-Z",
    journal = "Nucl. Phys. B",
    volume = "343",
    pages = "1--13",
    year = "1990"
}

@article{Sterman:1997sx,
    author = "Sterman, George F. and Stoler, Paul",
    title = "{Hadronic form-factors and perturbative QCD}",
    eprint = "hep-ph/9708370",
    archivePrefix = "arXiv",
    reportNumber = "ITP-SB-97-49, RPI-SP-97-6",
    doi = "10.1146/annurev.nucl.47.1.193",
    journal = "Ann. Rev. Nucl. Part. Sci.",
    volume = "47",
    pages = "193--233",
    year = "1997"
}

@article{Charles:1998dr,
    author = "Charles, J. and Le Yaouanc, A. and Oliver, L. and Pene, O. and Raynal, J. C.",
    title = "{Heavy to light form-factors in the heavy mass to large energy limit of QCD}",
    eprint = "hep-ph/9812358",
    archivePrefix = "arXiv",
    reportNumber = "LPTHE-ORSAY-98-77",
    doi = "10.1103/PhysRevD.60.014001",
    journal = "Phys. Rev. D",
    volume = "60",
    pages = "014001",
    year = "1999"
}

@article{Soares:1996vs,
    author = "Soares, Joao M.",
    title = "{Form-factor relations for heavy to heavy and heavy to light meson transitions}",
    eprint = "hep-ph/9607284",
    archivePrefix = "arXiv",
    reportNumber = "UMHEP-430",
    doi = "10.1103/PhysRevD.54.6837",
    journal = "Phys. Rev. D",
    volume = "54",
    pages = "6837--6841",
    year = "1996"
}

@article{Stech:1995ec,
    author = "Stech, Berthold",
    title = "{Form-factor relations for heavy to light transitions}",
    eprint = "hep-ph/9502378",
    archivePrefix = "arXiv",
    reportNumber = "HD-THEP-95-4",
    doi = "10.1016/0370-2693(95)00661-4",
    journal = "Phys. Lett. B",
    volume = "354",
    pages = "447--452",
    year = "1995"
}

@article{Shifman:1978bx,
    author = "Shifman, Mikhail A. and Vainshtein, A. I. and Zakharov, Valentin I.",
    title = "{QCD and Resonance Physics. Theoretical Foundations}",
    reportNumber = "ITEP-73-1978, ITEP-80-1978",
    doi = "10.1016/0550-3213(79)90022-1",
    journal = "Nucl. Phys. B",
    volume = "147",
    pages = "385--447",
    year = "1979"
}

@article{Shifman:1978by,
    author = "Shifman, Mikhail A. and Vainshtein, A. I. and Zakharov, Valentin I.",
    title = "{QCD and Resonance Physics: Applications}",
    reportNumber = "ITEP-94-1978, ITEP-81-1978",
    doi = "10.1016/0550-3213(79)90023-3",
    journal = "Nucl. Phys. B",
    volume = "147",
    pages = "448--518",
    year = "1979"
}

@article{Reinders:1984sr,
    author = "Reinders, L. J. and Rubinstein, H. and Yazaki, S.",
    title = "{Hadron Properties from QCD Sum Rules}",
    reportNumber = "CERN-TH-4079-84",
    doi = "10.1016/0370-1573(85)90065-1",
    journal = "Phys. Rept.",
    volume = "127",
    pages = "1",
    year = "1985"
}

@book{Narison:1989aq,
    author = "Narison, Stephan",
    title = "{QCD spectral sum rules}",
    volume = "26",
    year = "1989"
}

@book{Khodjamirian:2020btr,
    author = "Khodjamirian, Alexander",
    title = "{Hadron Form Factors}",
    doi = "10.1201/9781315142005",
    isbn = "978-1-138-30675-2, 978-1-315-14200-5",
    publisher = "CRC Press",
    month = "4",
    year = "2020"
}

@article{Colangelo:2000dp,
    author = "Colangelo, Pietro and Khodjamirian, Alexander",
    editor = "Shifman, M. and Ioffe, Boris",
    title = "{QCD sum rules, a modern perspective}",
    eprint = "hep-ph/0010175",
    archivePrefix = "arXiv",
    reportNumber = "CERN-TH-2000-296, BARI-TH-2000-394",
    doi = "10.1142/9789812810458_0033",
    pages = "1495--1576",
    month = "10",
    year = "2000"
}

@article{Ball:2004ye,
    author = "Ball, Patricia and Zwicky, Roman",
    title = "{New results on $B \to \pi, K, \eta$ decay formfactors from light-cone sum rules}",
    eprint = "hep-ph/0406232",
    archivePrefix = "arXiv",
    reportNumber = "IPPP-04-23, DCPT-04-46, TPI-MINN-04-25",
    doi = "10.1103/PhysRevD.71.014015",
    journal = "Phys. Rev. D",
    volume = "71",
    pages = "014015",
    year = "2005"
}

@article{Ball:2004rg,
    author = "Ball, Patricia and Zwicky, Roman",
    title = "{$B_{d,s} \to  \rho, \omega, K^*, \phi$ decay form-factors from light-cone sum rules revisited}",
    eprint = "hep-ph/0412079",
    archivePrefix = "arXiv",
    reportNumber = "IPPP-04-74, DCPT-04-48, TPI-MINN-04-39",
    doi = "10.1103/PhysRevD.71.014029",
    journal = "Phys. Rev. D",
    volume = "71",
    pages = "014029",
    year = "2005"
}

@article{Kruger:1999xa,
    author = "Kruger, Frank and Sehgal, Lalit M. and Sinha, Nita and Sinha, Rahul",
    title = "{Angular distribution and CP asymmetries in the decays $\bar B \to K^- \pi^+ e^- e^+$ and $\bar B \to \pi^- \pi^+ e^- e^+$}",
    eprint = "hep-ph/9907386",
    archivePrefix = "arXiv",
    reportNumber = "FISIST-5-99-CFIF, IMSC-99-05-17, PITHA-99-15",
    doi = "10.1103/PhysRevD.61.114028",
    journal = "Phys. Rev. D",
    volume = "61",
    pages = "114028",
    year = "2000",
    note = "[Erratum: Phys.Rev.D 63, 019901 (2001)]"
}

@article{Altmannshofer:2008dz,
    author = "Altmannshofer, Wolfgang and Ball, Patricia and Bharucha, Aoife and Buras, Andrzej J. and Straub, David M. and Wick, Michael",
    title = "{Symmetries and Asymmetries of $B \to K^{*} \mu^{+} \mu^{-}$ Decays in the Standard Model and Beyond}",
    eprint = "0811.1214",
    archivePrefix = "arXiv",
    primaryClass = "hep-ph",
    reportNumber = "IPPP-08-58, DCPT-08-116, TUM-HEP-696-08",
    doi = "10.1088/1126-6708/2009/01/019",
    journal = "JHEP",
    volume = "01",
    pages = "019",
    year = "2009"
}

@article{Descotes-Genon:2012isb,
    author = "Descotes-Genon, Sebastien and Matias, Joaquim and Ramon, Marc and Virto, Javier",
    title = "{Implications from clean observables for the binned analysis of $B -> K*\mu^+\mu^-$ at large recoil}",
    eprint = "1207.2753",
    archivePrefix = "arXiv",
    primaryClass = "hep-ph",
    reportNumber = "UAB-FT-275, LPT-ORSAY-12-81, UAB-FT-710",
    doi = "10.1007/JHEP01(2013)048",
    journal = "JHEP",
    volume = "01",
    pages = "048",
    year = "2013"
}

@article{Matias:2012xw,
    author = "Matias, Joaquim and Mescia, Federico and Ramon, Marc and Virto, Javier",
    title = "{Complete Anatomy of $\bar{B}_d -> \bar{K}^{* 0} (-> K \pi)l^+l^-$ and its angular distribution}",
    eprint = "1202.4266",
    archivePrefix = "arXiv",
    primaryClass = "hep-ph",
    reportNumber = "UAB-FT-706, ICCUB-12-076, ECM-UB-68",
    doi = "10.1007/JHEP04(2012)104",
    journal = "JHEP",
    volume = "04",
    pages = "104",
    year = "2012"
}

@article{Egede:2008uy,
    author = "Egede, U. and Hurth, T. and Matias, J. and Ramon, M. and Reece, W.",
    title = "{New observables in the decay mode $\bar B_d \to \bar K^{*0} l^+ l^-$}",
    eprint = "0807.2589",
    archivePrefix = "arXiv",
    primaryClass = "hep-ph",
    reportNumber = "SLAC-PUB-13307, UAB-FT-2008-652, CERN-TH-2008-155",
    doi = "10.1088/1126-6708/2008/11/032",
    journal = "JHEP",
    volume = "11",
    pages = "032",
    year = "2008"
}

@article{Hiller:2003js,
    author = "Hiller, Gudrun and Kruger, Frank",
    title = "{More model-independent analysis of $b \to s$ processes}",
    eprint = "hep-ph/0310219",
    archivePrefix = "arXiv",
    reportNumber = "LMU-18-03, TUM-HEP-519-03",
    doi = "10.1103/PhysRevD.69.074020",
    journal = "Phys. Rev. D",
    volume = "69",
    pages = "074020",
    year = "2004"
}

@article{Hiller:2014yaa,
    author = "Hiller, Gudrun and Schmaltz, Martin",
    title = "{$R_K$ and future $b \to s \ell \ell$ physics beyond the standard model opportunities}",
    eprint = "1408.1627",
    archivePrefix = "arXiv",
    primaryClass = "hep-ph",
    reportNumber = "DO-TH-14-17",
    doi = "10.1103/PhysRevD.90.054014",
    journal = "Phys. Rev. D",
    volume = "90",
    pages = "054014",
    year = "2014"
}

@article{Chobanova:2017ghn,
    author = "Chobanova, V. G. and Hurth, T. and Mahmoudi, F. and Martinez Santos, D. and Neshatpour, S.",
    title = "{Large hadronic power corrections or new physics in the rare decay $B→K^{\ast}\mu^{+}\mu^{−}$?}",
    eprint = "1702.02234",
    archivePrefix = "arXiv",
    primaryClass = "hep-ph",
    reportNumber = "CERN-TH-2017-031, IPM-PA-455, MITP-17-007",
    doi = "10.1007/JHEP07(2017)025",
    journal = "JHEP",
    volume = "07",
    pages = "025",
    year = "2017"
}

@article{Jager:2012uw,
    author = {J{\"a}ger, S. and Martin Camalich, J.},
    title = "{On $B \to  V \ell \ell$ at small dilepton invariant mass, power corrections, and new physics}",
    eprint = "1212.2263",
    archivePrefix = "arXiv",
    primaryClass = "hep-ph",
    doi = "10.1007/JHEP05(2013)043",
    journal = "JHEP",
    volume = "05",
    pages = "043",
    year = "2013"
}

@article{Descotes-Genon:2014uoa,
    author = "Descotes-Genon, S{\'e}bastien and Hofer, Lars and Matias, Joaquim and Virto, Javier",
    title = "{On the impact of power corrections in the prediction of $B \to K^*\mu^+\mu^-$ observables}",
    eprint = "1407.8526",
    archivePrefix = "arXiv",
    primaryClass = "hep-ph",
    reportNumber = "LPT-ORSAY-14-63, UAB-FT-757, QFET-2014-13, SI-HEP-2014-19",
    doi = "10.1007/JHEP12(2014)125",
    journal = "JHEP",
    volume = "12",
    pages = "125",
    year = "2014"
}

@article{Jager:2014rwa,
    author = {J{\"a}ger, Sebastian and Martin Camalich, Jorge},
    title = "{Reassessing the discovery potential of the $B \to K^{*} \ell^+\ell^-$ decays in the large-recoil region: SM challenges and BSM opportunities}",
    eprint = "1412.3183",
    archivePrefix = "arXiv",
    primaryClass = "hep-ph",
    doi = "10.1103/PhysRevD.93.014028",
    journal = "Phys. Rev. D",
    volume = "93",
    number = "1",
    pages = "014028",
    year = "2016"
}

@article{Ciuchini:2020gvn,
    author = "Ciuchini, Marco and Fedele, Marco and Franco, Enrico and Paul, Ayan and Silvestrini, Luca and Valli, Mauro",
    title = "{Lessons from the $B^{0,+}\to K^{*0,+}\mu^+\mu^-$ angular analyses}",
    eprint = "2011.01212",
    archivePrefix = "arXiv",
    primaryClass = "hep-ph",
    reportNumber = "DESY 20-190, DESY-20-190, HU-EP-20/30, HU 20/30, TTP20-037, P3H-20-064, UCI-TR 2020-18",
    doi = "10.1103/PhysRevD.103.015030",
    journal = "Phys. Rev. D",
    volume = "103",
    number = "1",
    pages = "015030",
    year = "2021"
}

@article{Altmannshofer:2026cwk,
    author = "Altmannshofer, Wolfgang and Christensen, Samuel G. and Stangl, Peter",
    title = "{Large Hadronic Effects in $B \to K^* \mu\mu$?}",
    eprint = "2603.27753",
    archivePrefix = "arXiv",
    primaryClass = "hep-ph",
    reportNumber = "MITP-26-013",
    month = "3",
    year = "2026"
}

@article{Descotes-Genon:2013vna,
    author = "Descotes-Genon, Sebastien and Hurth, Tobias and Matias, Joaquim and Virto, Javier",
    title = "{Optimizing the basis of $B\to K^*ll$ observables in the full kinematic range}",
    eprint = "1303.5794",
    archivePrefix = "arXiv",
    primaryClass = "hep-ph",
    reportNumber = "UAB-FT-732, LPT-ORSAY-13-24, MITP-13-020",
    doi = "10.1007/JHEP05(2013)137",
    journal = "JHEP",
    volume = "05",
    pages = "137",
    year = "2013"
}

@article{Ciuchini:1997hb,
    author = "Ciuchini, Marco and Franco, E. and Martinelli, G. and Silvestrini, L.",
    title = "{Charming penguins in B decays}",
    eprint = "hep-ph/9703353",
    archivePrefix = "arXiv",
    reportNumber = "CERN-TH-97-030, CERN-TH-97-30, ROME-97-1168, ROME-1168-1997, ROM2F-97-09, ROM2F--97-09",
    doi = "10.1016/S0550-3213(97)00388-X",
    journal = "Nucl. Phys. B",
    volume = "501",
    pages = "271--296",
    year = "1997"
}

@article{Bauer:2004tj,
    author = "Bauer, Christian W. and Pirjol, Dan and Rothstein, Ira Z. and Stewart, Iain W.",
    title = "{B ---{\ensuremath{>}} M(1) M(2): Factorization, charming penguins, strong phases, and polarization}",
    eprint = "hep-ph/0401188",
    archivePrefix = "arXiv",
    reportNumber = "MIT-CTP-3469, CMU-HEP-04-01, CALT-68-2475",
    doi = "10.1103/PhysRevD.70.054015",
    journal = "Phys. Rev. D",
    volume = "70",
    pages = "054015",
    year = "2004"
}

@article{Ciuchini:2015qxb,
    author = "Ciuchini, Marco and Fedele, Marco and Franco, Enrico and Mishima, Satoshi and Paul, Ayan and Silvestrini, Luca and Valli, Mauro",
    title = "{$B\to K^* \ell^+ \ell^-$ decays at large recoil in the Standard Model: a theoretical reappraisal}",
    eprint = "1512.07157",
    archivePrefix = "arXiv",
    primaryClass = "hep-ph",
    doi = "10.1007/JHEP06(2016)116",
    journal = "JHEP",
    volume = "06",
    pages = "116",
    year = "2016"
}

@article{Ciuchini:2021smi,
    author = "Ciuchini, Marco and Fedele, Marco and Franco, Enrico and Paul, Ayan and Silvestrini, Luca and Valli, Mauro",
    title = "{Charming penguins and lepton universality violation in~$b \to s \ell ^+ \ell ^-$~decays}",
    eprint = "2110.10126",
    archivePrefix = "arXiv",
    primaryClass = "hep-ph",
    reportNumber = "DESY 21-168, HU-EP-21/42, TTP21-040, P3H-21-077, YITP-SB-2021-20",
    doi = "10.1140/epjc/s10052-023-11191-w",
    journal = "Eur. Phys. J. C",
    volume = "83",
    number = "1",
    pages = "64",
    year = "2023"
}

@article{Khodjamirian:2010vf,
    author = "Khodjamirian, A. and Mannel, Th. and Pivovarov, A. A. and Wang, Y. -M.",
    title = "{Charm-loop effect in $B \to K^{(*)} \ell^{+} \ell^{-}$ and $B\to K^*\gamma$}",
    eprint = "1006.4945",
    archivePrefix = "arXiv",
    primaryClass = "hep-ph",
    reportNumber = "SI-HEP-2010-08",
    doi = "10.1007/JHEP09(2010)089",
    journal = "JHEP",
    volume = "09",
    pages = "089",
    year = "2010"
}

@article{Lyon:2014hpa,
    author = "Lyon, James and Zwicky, Roman",
    title = "{Resonances gone topsy turvy - the charm of QCD or new physics in $b \to s \ell^+ \ell^-$?}",
    eprint = "1406.0566",
    archivePrefix = "arXiv",
    primaryClass = "hep-ph",
    reportNumber = "EDINBURGH-14-10, CP3-ORIGINS-2014-021, DIAS-2014-21",
    month = "6",
    year = "2014"
}

@article{Gubernari:2020eft,
    author = "Gubernari, Nico and van Dyk, Danny and Virto, Javier",
    title = "{Non-local matrix elements in $B_{(s)}\to \{K^{(*)},\phi\}\ell^+\ell^-$}",
    eprint = "2011.09813",
    archivePrefix = "arXiv",
    primaryClass = "hep-ph",
    reportNumber = "EOS-2020-01, TUM-HEP-1292/20, P3H-20-066, SI-HEP-2020-27",
    doi = "10.1007/JHEP02(2021)088",
    journal = "JHEP",
    volume = "02",
    pages = "088",
    year = "2021"
}

@article{Mahajan:2024xpo,
    author = "Mahajan, Namit and Mishra, Dayanand",
    title = "{Smallness of charm-loop effects in B{\textrightarrow}K(*){\ensuremath{\ell}}{\ensuremath{\ell}} at low q2: Light-meson distribution-amplitude analysis}",
    eprint = "2409.00181",
    archivePrefix = "arXiv",
    primaryClass = "hep-ph",
    doi = "10.1103/PhysRevD.111.L031504",
    journal = "Phys. Rev. D",
    volume = "111",
    number = "3",
    pages = "L031504",
    year = "2025"
}

@article{Isidori:2024lng,
    author = "Isidori, Gino and Polonsky, Zachary and Tinari, Arianna",
    title = "{Explicit estimate of charm rescattering in B0{\textrightarrow}K0{\ensuremath{\ell}}{\textasciimacron}{\ensuremath{\ell}}}",
    eprint = "2405.17551",
    archivePrefix = "arXiv",
    primaryClass = "hep-ph",
    doi = "10.1103/PhysRevD.111.093007",
    journal = "Phys. Rev. D",
    volume = "111",
    number = "9",
    pages = "093007",
    year = "2025"
}

@article{Atwood:1997bn,
    author = "Atwood, David and Soni, Amarjit",
    title = "{B ---{\ensuremath{>}} eta-prime + X and the QCD anomaly}",
    eprint = "hep-ph/9704357",
    archivePrefix = "arXiv",
    reportNumber = "JLAB-THY-97-18",
    doi = "10.1016/S0370-2693(97)00592-3",
    journal = "Phys. Lett. B",
    volume = "405",
    pages = "150--156",
    year = "1997"
}

@article{Hou:1997wy,
    author = "Hou, Wei-Shu and Tseng, B.",
    title = "{Enhanced b ---{\ensuremath{>}} s g decay, inclusive eta-prime production, and the gluon anomaly}",
    eprint = "hep-ph/9705304",
    archivePrefix = "arXiv",
    doi = "10.1103/PhysRevLett.80.434",
    journal = "Phys. Rev. Lett.",
    volume = "80",
    pages = "434--437",
    year = "1998"
}

@article{Halperin:1997ma,
    author = "Halperin, Igor E. and Zhitnitsky, Ariel",
    title = "{Why is the B ---{\ensuremath{>}} eta-prime X decay width so large?}",
    eprint = "hep-ph/9705251",
    archivePrefix = "arXiv",
    doi = "10.1103/PhysRevLett.80.438",
    journal = "Phys. Rev. Lett.",
    volume = "80",
    pages = "438--441",
    year = "1998"
}

@article{Halperin:1997as,
    author = "Halperin, Igor E. and Zhitnitsky, Ariel",
    title = "{B ---{\ensuremath{>}} K eta-prime decay as unique probe of eta-prime meson}",
    eprint = "hep-ph/9704412",
    archivePrefix = "arXiv",
    doi = "10.1103/PhysRevD.56.7247",
    journal = "Phys. Rev. D",
    volume = "56",
    pages = "7247--7258",
    year = "1997"
}

@article{Shuryak:1997xd,
    author = "Shuryak, Edward V. and Zhitnitsky, A. R.",
    title = "{The Gluon / charm content of the eta-prime meson and instantons}",
    eprint = "hep-ph/9706316",
    archivePrefix = "arXiv",
    reportNumber = "NI-97033-NQF",
    doi = "10.1103/PhysRevD.57.2001",
    journal = "Phys. Rev. D",
    volume = "57",
    pages = "2001--2004",
    year = "1998"
}

@article{Yuan:1997ts,
    author = "Yuan, Feng and Chao, Kuang-Ta",
    title = "{The Color octet intrinsic charm in eta-prime and b ---{\ensuremath{>}} eta-prime X decays}",
    eprint = "hep-ph/9706294",
    archivePrefix = "arXiv",
    reportNumber = "PKU-TP-97-21",
    doi = "10.1103/PhysRevD.56.R2495",
    journal = "Phys. Rev. D",
    volume = "56",
    pages = "R2495--R2498",
    year = "1997"
}

@article{Ali:1997nh,
    author = "Ali, Ahmed and Greub, C.",
    title = "{An Analysis of two-body nonleptonic B decays involving light mesons in the standard model}",
    eprint = "hep-ph/9707251",
    archivePrefix = "arXiv",
    reportNumber = "DESY-97-126",
    doi = "10.1103/PhysRevD.57.2996",
    journal = "Phys. Rev. D",
    volume = "57",
    pages = "2996--3016",
    year = "1998"
}

@article{Chen:1999nxa,
    author = "Chen, Yaw-Hwang and Cheng, Hai-Yang and Tseng, B. and Yang, Kwei-Chou",
    title = "{Charmless hadronic two-body decays of B(u) and B(d) mesons}",
    eprint = "hep-ph/9903453",
    archivePrefix = "arXiv",
    reportNumber = "IP-ASTP-01-99",
    doi = "10.1103/PhysRevD.60.094014",
    journal = "Phys. Rev. D",
    volume = "60",
    pages = "094014",
    year = "1999"
}

@article{Ali:1997ex,
    author = "Ali, Ahmed and Chay, J. and Greub, C. and Ko, P.",
    title = "{Contribution of b ---{\ensuremath{>}} s gluon gluon through the QCD anomaly in exclusive decays B+- ---{\ensuremath{>}} (eta-prime, eta) (K+-, K*+-) and B0 ---{\ensuremath{>}} (eta-prime, eta) (K0, K*0)}",
    eprint = "hep-ph/9712372",
    archivePrefix = "arXiv",
    reportNumber = "DESY-97-235, KAIST-TH-12-97, BUTP-97-34, SNUTP-97-169",
    doi = "10.1016/S0370-2693(98)00174-9",
    journal = "Phys. Lett. B",
    volume = "424",
    pages = "161--174",
    year = "1998"
}

@article{Bagchi:1999dx,
    author = "Bagchi, B. and Bhattacharyya, P. and Sen, S. and Chakrabarti, J.",
    title = "{Mixing angles and decay constants of eta, eta-prime and eta(c)}",
    doi = "10.1103/PhysRevD.60.074002",
    journal = "Phys. Rev. D",
    volume = "60",
    pages = "074002",
    year = "1999"
}

@article{Fritzsch:1997ps,
    author = "Fritzsch, Harald",
    title = "{The Gluonic decay of the b quark and the eta-prime meson}",
    eprint = "hep-ph/9708348",
    archivePrefix = "arXiv",
    reportNumber = "CERN-TH-97-200, LMU-08-97",
    doi = "10.1016/S0370-2693(97)01213-6",
    journal = "Phys. Lett. B",
    volume = "415",
    pages = "83--89",
    year = "1997"
}

@article{Chao:1989yp,
    author = "Chao, Kuang-Ta",
    title = "{{QCD} Axial Anomaly and Mixing of Pseudoscalars}",
    doi = "10.1016/0550-3213(89)90534-8",
    journal = "Nucl. Phys. B",
    volume = "317",
    pages = "597--616",
    year = "1989"
}

@article{Chao:1990im,
    author = "Chao, Kuang-Ta",
    title = "{Mixing of eta, eta-prime with c anti-c, b anti-b states and their radiative decays}",
    doi = "10.1016/0550-3213(90)90172-A",
    journal = "Nucl. Phys. B",
    volume = "335",
    pages = "101--114",
    year = "1990"
}

@article{Feldmann:1998vh,
    author = "Feldmann, T. and Kroll, P. and Stech, B.",
    title = "{Mixing and decay constants of pseudoscalar mesons}",
    eprint = "hep-ph/9802409",
    archivePrefix = "arXiv",
    reportNumber = "WU-B-98-2, HD-THEP-98-5",
    doi = "10.1103/PhysRevD.58.114006",
    journal = "Phys. Rev. D",
    volume = "58",
    pages = "114006",
    year = "1998"
}

@article{Bodwin:1994jh,
    author = "Bodwin, Geoffrey T. and Braaten, Eric and Lepage, G. Peter",
    title = "{Rigorous QCD analysis of inclusive annihilation and production of heavy quarkonium}",
    eprint = "hep-ph/9407339",
    archivePrefix = "arXiv",
    reportNumber = "ANL-HEP-PR-94-24, FERMILAB-PUB-94-073-T, NUHEP-TH-94-5",
    doi = "10.1103/PhysRevD.55.5853",
    journal = "Phys. Rev. D",
    volume = "51",
    pages = "1125--1171",
    year = "1995",
    note = "[Erratum: Phys.Rev.D 55, 5853 (1997)]"
}

@article{Kiselev:1992tx,
    author = "Kiselev, V. V. and Likhoded, A. K. and Tkabladze, A. V.",
    title = "{Semileptonic B(c) decays}",
    reportNumber = "IFVE-92-138",
    journal = "Phys. Atom. Nucl.",
    volume = "56",
    pages = "643--649",
    year = "1993"
}

@article{Kiselev:1992au,
    author = "Kiselev, V. V.",
    title = "{Leptonic decay constants of heavy quarkonia in effective QCD sum rules}",
    reportNumber = "IFVE-92-149",
    doi = "10.1016/0550-3213(93)90171-K",
    journal = "Nucl. Phys. B",
    volume = "406",
    pages = "340--354",
    year = "1993"
}

@article{Kiselev:1995bv,
    author = "Kiselev, V. V.",
    title = "{Scaling relations in phenomenology of QCD sum rules for heavy quarkonium}",
    eprint = "hep-ph/9504313",
    archivePrefix = "arXiv",
    reportNumber = "IFVE-95-63, IHEP-95-63",
    doi = "10.1142/S0217751X96001723",
    journal = "Int. J. Mod. Phys. A",
    volume = "11",
    pages = "3689--3710",
    year = "1996"
}

@article{Jenkins:1992nb,
    author = "Jenkins, Elizabeth Ellen and Luke, Michael E. and Manohar, Aneesh V. and Savage, Martin J.",
    title = "{Semileptonic B(c) decay and heavy quark spin symmetry}",
    eprint = "hep-ph/9204238",
    archivePrefix = "arXiv",
    reportNumber = "UCSD-PTH-92-13",
    doi = "10.1016/0550-3213(93)90464-Z",
    journal = "Nucl. Phys. B",
    volume = "390",
    pages = "463--473",
    year = "1993"
}

@article{Wise:1992hn,
    author = "Wise, Mark B.",
    title = "{Chiral perturbation theory for hadrons containing a heavy quark}",
    reportNumber = "CALT-68-1765",
    doi = "10.1103/PhysRevD.45.R2188",
    journal = "Phys. Rev. D",
    volume = "45",
    number = "7",
    pages = "R2188",
    year = "1992"
}

@article{Yan:1992gz,
    author = "Yan, Tung-Mow and Cheng, Hai-Yang and Cheung, Chi-Yee and Lin, Guey-Lin and Lin, Y. C. and Yu, Hoi-Lai",
    title = "{Heavy quark symmetry and chiral dynamics}",
    reportNumber = "CLNS-92-1138, IP-ASTP-03-92",
    doi = "10.1103/PhysRevD.46.1148",
    journal = "Phys. Rev. D",
    volume = "46",
    pages = "1148--1164",
    year = "1992",
    note = "[Erratum: Phys.Rev.D 55, 5851 (1997)]"
}

@article{Burdman:1992gh,
    author = "Burdman, Gustavo and Donoghue, John F.",
    title = "{Union of chiral and heavy quark symmetries}",
    reportNumber = "UMHEP-365",
    doi = "10.1016/0370-2693(92)90068-F",
    journal = "Phys. Lett. B",
    volume = "280",
    pages = "287--291",
    year = "1992"
}

@article{Cho:1992gg,
    author = "Cho, Peter L.",
    title = "{Chiral perturbation theory for hadrons containing a heavy quark: The Sequel}",
    eprint = "hep-ph/9203225",
    archivePrefix = "arXiv",
    reportNumber = "HUTP-92-A014",
    doi = "10.1016/0370-2693(92)91314-Y",
    journal = "Phys. Lett. B",
    volume = "285",
    pages = "145--152",
    year = "1992"
}

@article{Cheng:1992xi,
    author = "Cheng, Hai-Yang and Cheung, Chi-Yee and Lin, Guey-Lin and Lin, Y. C. and Yan, Tung-Mow and Yu, Hoi-Lai",
    title = "{Chiral Lagrangians for radiative decays of heavy hadrons}",
    eprint = "hep-ph/9209262",
    archivePrefix = "arXiv",
    reportNumber = "CLNS-92-1158, IP-ASTP-13-92",
    doi = "10.1103/PhysRevD.47.1030",
    journal = "Phys. Rev. D",
    volume = "47",
    pages = "1030--1042",
    year = "1993"
}

@article{Grinstein:1992qt,
    author = "Grinstein, Benjamin and Jenkins, Elizabeth Ellen and Manohar, Aneesh V. and Savage, Martin J. and Wise, Mark B.",
    title = "{Chiral perturbation theory for f D(s) / f D and B B(s) / B B}",
    eprint = "hep-ph/9204207",
    archivePrefix = "arXiv",
    reportNumber = "UCSD-PTH-92-05, CALT-68-1768, SSCL-PREPRINT-025",
    doi = "10.1016/0550-3213(92)90248-A",
    journal = "Nucl. Phys. B",
    volume = "380",
    pages = "369--376",
    year = "1992"
}

@article{Cho:1992cf,
    author = "Cho, Peter L.",
    title = "{Heavy hadron chiral perturbation theory}",
    eprint = "hep-ph/9208244",
    archivePrefix = "arXiv",
    reportNumber = "HUTP-92-A039",
    doi = "10.1016/0550-3213(93)90263-O",
    journal = "Nucl. Phys. B",
    volume = "396",
    pages = "183--204",
    year = "1993",
    note = "[Erratum: Nucl.Phys.B 421, 683--686 (1994)]"
}

@article{Casalbuoni:1996pg,
    author = "Casalbuoni, R. and Deandrea, A. and Di Bartolomeo, N. and Gatto, Raoul and Feruglio, F. and Nardulli, G.",
    title = "{Phenomenology of heavy meson chiral Lagrangians}",
    eprint = "hep-ph/9605342",
    archivePrefix = "arXiv",
    reportNumber = "UGVA-DPT-1996-05-928, BARI-TH-96-237",
    doi = "10.1016/S0370-1573(96)00027-0",
    journal = "Phys. Rept.",
    volume = "281",
    pages = "145--238",
    year = "1997"
}

@article{Falk:1993fr,
    author = "Falk, Adam F. and Grinstein, Benjamin",
    title = "{Anti-B ---{\ensuremath{>}} Anti-K e+ e- in Chiral Perturbation Theory}",
    eprint = "hep-ph/9306310",
    archivePrefix = "arXiv",
    reportNumber = "SLAC-PUB-6237, SSCL-PREPRINT-484",
    doi = "10.1016/0550-3213(94)90554-1",
    journal = "Nucl. Phys. B",
    volume = "416",
    pages = "771--785",
    year = "1994"
}

@article{Bardeen:1993ae,
    author = "Bardeen, William A. and Hill, Christopher T.",
    title = "{Chiral dynamics and heavy quark symmetry in a solvable toy field theoretic model}",
    eprint = "hep-ph/9304265",
    archivePrefix = "arXiv",
    reportNumber = "SSCL-PREPRINT-243, SSCL-PP-243, FERMILAB-PUB-93-059-T",
    doi = "10.1103/PhysRevD.49.409",
    journal = "Phys. Rev. D",
    volume = "49",
    pages = "409--425",
    year = "1994"
}

@article{Nowak:1992um,
    author = "Nowak, Maciej A. and Rho, Mannque and Zahed, I.",
    title = "{Chiral effective action with heavy quark symmetry}",
    eprint = "hep-ph/9209272",
    archivePrefix = "arXiv",
    reportNumber = "SUNY-NTG-92-27",
    doi = "10.1103/PhysRevD.48.4370",
    journal = "Phys. Rev. D",
    volume = "48",
    pages = "4370--4374",
    year = "1993"
}

@article{Ebert:1994tv,
    author = "Ebert, D. and Feldmann, T. and Friedrich, R. and Reinhardt, H.",
    title = "{Effective meson Lagrangian with chiral and heavy quark symmetries from quark flavor dynamics}",
    eprint = "hep-ph/9406220",
    archivePrefix = "arXiv",
    reportNumber = "HUB-IEP-94-8, UNITU-THEP-11-94, DESY-94-098",
    doi = "10.1016/0550-3213(94)00456-O",
    journal = "Nucl. Phys. B",
    volume = "434",
    pages = "619--646",
    year = "1995"
}

@article{Beneke:1999br,
    author = "Beneke, M. and Buchalla, G. and Neubert, M. and Sachrajda, Christopher T.",
    title = "{QCD factorization for B ---{\ensuremath{>}} pi pi decays: Strong phases and CP violation in the heavy quark limit}",
    eprint = "hep-ph/9905312",
    archivePrefix = "arXiv",
    reportNumber = "SLAC-PUB-8146, CERN-TH-99-126, SHEP-99-04",
    doi = "10.1103/PhysRevLett.83.1914",
    journal = "Phys. Rev. Lett.",
    volume = "83",
    pages = "1914--1917",
    year = "1999"
}

@article{Beneke:2001ev,
    author = "Beneke, M. and Buchalla, G. and Neubert, M. and Sachrajda, Christopher T.",
    title = "{QCD factorization in B ---{\ensuremath{>}} pi K, pi pi decays and extraction of Wolfenstein parameters}",
    eprint = "hep-ph/0104110",
    archivePrefix = "arXiv",
    reportNumber = "CERN-TH-2001-107, CLNS-01-1728, PITHA-01-01, SHEP-01-11",
    doi = "10.1016/S0550-3213(01)00251-6",
    journal = "Nucl. Phys. B",
    volume = "606",
    pages = "245--321",
    year = "2001"
}

@article{Keum:2000ms,
    author = "Keum, Yong-Yeon and Li, Hsiang-nan",
    title = "{Nonleptonic charmless B decays: Factorization versus perturbative QCD}",
    eprint = "hep-ph/0006001",
    archivePrefix = "arXiv",
    reportNumber = "NCKU-HEP-00-03, IPAS-HEP-2003, APCTP-00-09",
    doi = "10.1103/PhysRevD.63.074006",
    journal = "Phys. Rev. D",
    volume = "63",
    pages = "074006",
    year = "2001"
}

@article{Yeh:1997rq,
    author = "Yeh, Tsung-Wen and Li, Hsiang-nan",
    title = "{Factorization theorems, effective field theory, and nonleptonic heavy meson decays}",
    eprint = "hep-ph/9701233",
    archivePrefix = "arXiv",
    reportNumber = "CCUTH-96-07",
    doi = "10.1103/PhysRevD.56.1615",
    journal = "Phys. Rev. D",
    volume = "56",
    pages = "1615--1631",
    year = "1997"
}

@article{Li:1994iu,
    author = "Li, Hsiang-nan and Yu, Hoi-Lai",
    title = "{Perturbative QCD analysis of B meson decays}",
    eprint = "hep-ph/9411308",
    archivePrefix = "arXiv",
    reportNumber = "CCUTH-94-04, IP-ASTP-12-94",
    doi = "10.1103/PhysRevD.53.2480",
    journal = "Phys. Rev. D",
    volume = "53",
    pages = "2480--2490",
    year = "1996"
}

@article{Beneke:2001at,
    author = "Beneke, M. and Feldmann, T. and Seidel, D.",
    title = "{Systematic approach to exclusive $B \to  V l^+ l^-$, $V \gamma$ decays}",
    eprint = "hep-ph/0106067",
    archivePrefix = "arXiv",
    reportNumber = "PITHA-01-05",
    doi = "10.1016/S0550-3213(01)00366-2",
    journal = "Nucl. Phys. B",
    volume = "612",
    pages = "25--58",
    year = "2001"
}

@article{Bauer:2000yr,
    author = "Bauer, Christian W. and Fleming, Sean and Pirjol, Dan and Stewart, Iain W.",
    title = "{An Effective field theory for collinear and soft gluons: Heavy to light decays}",
    eprint = "hep-ph/0011336",
    archivePrefix = "arXiv",
    reportNumber = "UCSD-PTH-00-28",
    doi = "10.1103/PhysRevD.63.114020",
    journal = "Phys. Rev. D",
    volume = "63",
    pages = "114020",
    year = "2001"
}

@article{Bauer:2001yt,
    author = "Bauer, Christian W. and Pirjol, Dan and Stewart, Iain W.",
    title = "{Soft collinear factorization in effective field theory}",
    eprint = "hep-ph/0109045",
    archivePrefix = "arXiv",
    reportNumber = "UCSD-PTH-01-15",
    doi = "10.1103/PhysRevD.65.054022",
    journal = "Phys. Rev. D",
    volume = "65",
    pages = "054022",
    year = "2002"
}

@article{Zeppenfeld:1980ex,
    author = "Zeppenfeld, D.",
    title = "{SU(3) Relations for B Meson Decays}",
    reportNumber = "MPI-PAE-PTH-34-80",
    doi = "10.1007/BF01429835",
    journal = "Z. Phys. C",
    volume = "8",
    pages = "77",
    year = "1981"
}

@article{Savage:1989ub,
    author = "Savage, Martin J. and Wise, Mark B.",
    title = "{SU(3) Predictions for Nonleptonic B Meson Decays}",
    reportNumber = "CALT-68-1535",
    doi = "10.1103/PhysRevD.39.3346",
    journal = "Phys. Rev. D",
    volume = "39",
    pages = "3346",
    year = "1989",
    note = "[Erratum: Phys.Rev.D 40, 3127 (1989)]"
}

@article{Chau:1990ay,
    author = "Chau, Ling-Lie and Cheng, Hai-Yang and Sze, W. K. and Yao, Herng and Tseng, Benjamin",
    title = "{Charmless nonleptonic rare decays of $B$ mesons}",
    reportNumber = "IP-ASTP-08-90, UCDPHYS-PUB-08-90",
    doi = "10.1103/PhysRevD.43.2176",
    journal = "Phys. Rev. D",
    volume = "43",
    pages = "2176--2192",
    year = "1991",
    note = "[Erratum: Phys.Rev.D 58, 019902 (1998)]"
}

@article{Grinstein:1996us,
    author = "Grinstein, Benjamin and Lebed, Richard F.",
    title = "{SU(3) decomposition of two-body B decay amplitudes}",
    eprint = "hep-ph/9602218",
    archivePrefix = "arXiv",
    reportNumber = "UCSD-PTH-96-01",
    doi = "10.1103/PhysRevD.53.6344",
    journal = "Phys. Rev. D",
    volume = "53",
    pages = "6344--6360",
    year = "1996"
}

@article{Gronau:1994rj,
    author = "Gronau, Michael and Hernandez, Oscar F. and London, David and Rosner, Jonathan L.",
    title = "{Decays of B mesons to two light pseudoscalars}",
    eprint = "hep-ph/9404283",
    archivePrefix = "arXiv",
    reportNumber = "TECHNION-PH-94-8, UDEM-LPN-TH-94-193, EFI-94-12",
    doi = "10.1103/PhysRevD.50.4529",
    journal = "Phys. Rev. D",
    volume = "50",
    pages = "4529--4543",
    year = "1994"
}

@article{Gronau:1995hm,
    author = "Gronau, Michael and Hernandez, Oscar F. and London, David and Rosner, Jonathan L.",
    title = "{Broken SU(3) symmetry in two-body B decays}",
    eprint = "hep-ph/9504326",
    archivePrefix = "arXiv",
    reportNumber = "TECHNION-PH-95-10, UDEM-GPP-TH-95-24, EFI-95-09",
    doi = "10.1103/PhysRevD.52.6356",
    journal = "Phys. Rev. D",
    volume = "52",
    pages = "6356--6373",
    year = "1995"
}

@article{Gronau:1995hn,
    author = "Gronau, Michael and Hernandez, Oscar F. and London, David and Rosner, Jonathan L.",
    title = "{Electroweak penguins and two-body B decays}",
    eprint = "hep-ph/9504327",
    archivePrefix = "arXiv",
    reportNumber = "TECHNION-PH-95-11, UDEM-GPP-TH-95-25, EFI-95-15",
    doi = "10.1103/PhysRevD.52.6374",
    journal = "Phys. Rev. D",
    volume = "52",
    pages = "6374--6382",
    year = "1995"
}

@article{Gronau:1990ka,
    author = "Gronau, Michael and London, David",
    title = "{Isospin analysis of CP asymmetries in B decays}",
    reportNumber = "DESY-90-106-REV, UDEM-LPN-TH24-REV",
    doi = "10.1103/PhysRevLett.65.3381",
    journal = "Phys. Rev. Lett.",
    volume = "65",
    pages = "3381--3384",
    year = "1990"
}

@article{Nir:1991cu,
    author = "Nir, Yosef and Quinn, Helen R.",
    title = "{Measuring CKM parameters with CP asymmetry and isospin analysis in B ---{\ensuremath{>}} pi K}",
    reportNumber = "SLAC-PUB-5408, WIS-91-1-PH",
    doi = "10.1103/PhysRevLett.67.541",
    journal = "Phys. Rev. Lett.",
    volume = "67",
    pages = "541--544",
    year = "1991"
}

@article{Lipkin:1991st,
    author = "Lipkin, Harry J. and Nir, Yosef and Quinn, Helen R. and Snyder, A.",
    title = "{Penguin trapping with isospin analysis and CP asymmetries in B decays}",
    reportNumber = "SLAC-PUB-5445, WIS-91-5-PH",
    doi = "10.1103/PhysRevD.44.1454",
    journal = "Phys. Rev. D",
    volume = "44",
    pages = "1454--1460",
    year = "1991"
}

@article{Beneke:2000ry,
    author = "Beneke, M. and Buchalla, G. and Neubert, M. and Sachrajda, Christopher T.",
    title = "{QCD factorization for exclusive, nonleptonic B meson decays: General arguments and the case of heavy light final states}",
    eprint = "hep-ph/0006124",
    archivePrefix = "arXiv",
    reportNumber = "CERN-TH-2000-159, CLNS-00-1675, PITHA-00-06, SHEP-00-06",
    doi = "10.1016/S0550-3213(00)00559-9",
    journal = "Nucl. Phys. B",
    volume = "591",
    pages = "313--418",
    year = "2000"
}

@article{Chiang:2002tv,
    author = "Chiang, Cheng-Wei and Rosner, Jonathan L.",
    title = "{Final state phases in B ---{\ensuremath{>}} D pi, d* pi, and D(rho) decays}",
    eprint = "hep-ph/0212274",
    archivePrefix = "arXiv",
    reportNumber = "ANL-HEP-PR-02-113, EFI-02-46",
    doi = "10.1103/PhysRevD.67.074013",
    journal = "Phys. Rev. D",
    volume = "67",
    pages = "074013",
    year = "2003"
}

@article{Fayyazuddin:2004ac,
    author = "Fayyazuddin",
    title = "{Final state phases in B ---{\ensuremath{>}} D pi, anti-D pi decays and CP asymmetry}",
    eprint = "hep-ph/0402189",
    archivePrefix = "arXiv",
    reportNumber = "NCP-QAU-0218-2004",
    doi = "10.1103/PhysRevD.70.114018",
    journal = "Phys. Rev. D",
    volume = "70",
    pages = "114018",
    year = "2004"
}

@inproceedings{Pirjol:2004yf,
    author = "Pirjol, Dan",
    title = "{Factorization in color-suppressed anti-B ---{\ensuremath{>}} D(*) pi decays from the soft-collinear effective theory}",
    booktitle = "{32nd International Conference on High Energy Physics}",
    eprint = "hep-ph/0411124",
    archivePrefix = "arXiv",
    doi = "10.1142/9789812702227_0150",
    pages = "799--802",
    month = "11",
    year = "2004"
}

@article{Wang:2024tnx,
    author = "Wang, Di",
    title = "{Isospin sum rules for the nonleptonic B decays}",
    eprint = "2405.20698",
    archivePrefix = "arXiv",
    primaryClass = "hep-ph",
    doi = "10.1140/epjc/s10052-024-13349-6",
    journal = "Eur. Phys. J. C",
    volume = "84",
    number = "9",
    pages = "949",
    year = "2024"
}

@article{Okubo:1963fa,
    author = "Okubo, S.",
    title = "{Phi meson and unitary symmetry model}",
    doi = "10.1016/S0375-9601(63)92548-9",
    journal = "Phys. Lett.",
    volume = "5",
    pages = "165--168",
    year = "1963"
}

@inbook{Zweig:1964jf,
    author = "Zweig, G.",
    editor = "Lichtenberg, D. B. and Rosen, Simon Peter",
    title = "{An SU(3) model for strong interaction symmetry and its breaking. Version 2}",
    booktitle = "{DEVELOPMENTS IN THE QUARK THEORY OF HADRONS. VOL. 1. 1964 - 1978}",
    reportNumber = "CERN-TH-412, NP-14146, PRINT-64-170",
    doi = "10.17181/CERN-TH-412",
    pages = "22--101",
    month = "2",
    year = "1964"
}

@article{Iizuka:1966fk,
    author = "Iizuka, Jugoro",
    title = "{Systematics and phenomenology of meson family}",
    doi = "10.1143/PTPS.37.21",
    journal = "Prog. Theor. Phys. Suppl.",
    volume = "37",
    pages = "21--34",
    year = "1966"
}

@article{Murayama:2021xfj,
    author = "Murayama, Hitoshi",
    title = "{Some Exact Results in QCD-like Theories}",
    eprint = "2104.01179",
    archivePrefix = "arXiv",
    primaryClass = "hep-th",
    doi = "10.1103/PhysRevLett.126.251601",
    journal = "Phys. Rev. Lett.",
    volume = "126",
    number = "25",
    pages = "251601",
    year = "2021"
}

@article{Csaki:2022cyg,
    author = "Cs{\'a}ki, Csaba and Gomes, Andrew and Murayama, Hitoshi and Noether, Bea and Varier, Digvijay Roy and Telem, Ofri",
    title = "{Guide to anomaly-mediated supersymmetry-breaking QCD}",
    eprint = "2212.03260",
    archivePrefix = "arXiv",
    primaryClass = "hep-th",
    doi = "10.1103/PhysRevD.107.054015",
    journal = "Phys. Rev. D",
    volume = "107",
    number = "5",
    pages = "054015",
    year = "2023"
}

@article{Csaki:2025atr,
    author = "Cs{\'a}ki, Csaba and Roy, Tuhin S. and Ruhdorfer, Maximilian and Youn, Taewook",
    title = "{Dynamical Up-quark Mass Generation in QCD-like theories}",
    eprint = "2505.07953",
    archivePrefix = "arXiv",
    primaryClass = "hep-ph",
    month = "5",
    year = "2025"
}

@article{Kondo:2025njf,
    author = "Kondo, Dan and Murayama, Hitoshi and Noether, Bea",
    title = "{Near-SUSY to non-SUSY crossover}",
    eprint = "2505.18138",
    archivePrefix = "arXiv",
    primaryClass = "hep-th",
    doi = "10.1103/9mjb-bp8g",
    journal = "Phys. Rev. D",
    volume = "112",
    number = "11",
    pages = "114021",
    year = "2025"
}

@article{Csaki:2023yas,
    author = "Cs{\'a}ki, Csaba and Tito D'Agnolo, Raffaele and Gupta, Rick S. and Kuflik, Eric and Roy, Tuhin S. and Ruhdorfer, Maximilian",
    title = "{On the dynamical origin of the ${\ensuremath{\eta'}}$ potential and the axion mass}",
    eprint = "2307.04809",
    archivePrefix = "arXiv",
    primaryClass = "hep-ph",
    doi = "10.1007/JHEP10(2023)139",
    journal = "JHEP",
    volume = "10",
    pages = "139",
    year = "2023"
}

\end{document}